\documentclass[a4paper,11pt]{article}
\usepackage{amsmath}
\numberwithin{equation}{section}

\usepackage{graphicx,amssymb,amstext,amsmath}

\usepackage{color}
\usepackage{hyperref}

\newtheorem{propo}{Proposition}[section]

\newcommand{\bc}{\begin{center}}
\newcommand{\ec}{\end{center}}
\def\ba#1{\begin{array}{#1}\displaystyle}
\newcommand{\ea}{\end{array}}

\newcommand{\beq}{\begin{equation}}
\newcommand{\eeq}{\end{equation}}
\newcommand{\beqa}{\begin{eqnarray}}
\newcommand{\eeqa}{\end{eqnarray}}
\newcommand{\no}{\nonumber}
\newcommand{\n}{\nonumber\\}
\newcommand{\bi}{\begin{itemize}}
\newcommand{\ei}{\end{itemize}}
\def\mato#1{\left(\ba{#1}} 
\def\matf{\ea\right)}

\def\lt#1{\left#1}
\def\rt#1{\right#1}
\def\t#1{\tilde{#1}}

\def\b#1{\bar{#1}}
\def\frc#1#2{\frac{#1}{#2}}

\newcommand{\p}{\partial}

\newcommand{\vac}{{\rm vac}}
\newcommand{\bra}{\langle}
\newcommand{\ket}{\rangle}

\newcommand{\R}{{\mathbb{R}}}

\newcommand{\Or}{{\cal O}}

\newcommand{\ep}{\epsilon}
\newcommand{\varep}{\varepsilon}
\newcommand{\lam}{\lambda}
\newcommand{\om}{\omega}

\newcommand{\Tr}{{\rm Tr}}

\newcommand{\End}{{\rm End}}

\newcommand{\liou}{{\cal L}}

\newcommand{\lb}{{\bf b}}

\newcommand{\lU}{{\bf U}}
\newcommand{\lZ}{{\bf Z}}

\newcommand{\nl}{\mbox{\tiny ${\circ\atop\circ}$}}

\newcommand{\mH}{\mathcal{H}}

\newcommand{\T}{\mathcal{T}}

\newcommand{\rhoni}{\rho^{(n)}_{\text{I}}}
\newcommand{\rhonia}{\rho^{(n)}_{\text{I}_a}}
\newcommand{\rhonib}{\rho^{(n)}_{\text{I}_b}}
\newcommand{\rhond}{\rho^{(n)}_{\text{D}}}
\newcommand{\TD}{\T_{\text{Dirac}}}
\newcommand{\tTD}{\tilde\T_{\text{Dirac}}}
\newcommand{\td}{\text{D}}

\newcommand{\ltr}{{\Large\text{\Tr}}}

\newcommand{\twopair}{_{(\nu_1,\ep_1)(\nu_2,\ep_2)}}

\newcommand{\ttt}{\textit}
\newcommand{\tbf}{\textbf}
\newcommand{\tsl}{\textsl}

\begin{document}
 
\begin{titlepage}
\vspace{0.2cm}
\begin{center}

{\Large {\bf Mixed-state form factors of $U(1)$ twist fields in the Dirac theory}}

\vspace{0.8cm} {\large \text{Yixiong Chen}}

\vspace{0.2cm}
{Department of Mathematics, King's College London, Strand WC2R 2LS, UK }
\vspace{1cm} 
\end{center}

\begin{abstract}
\vspace{0.5cm} 
Using the ``Liouville space'' (the space of operators) of the massive Dirac theory, we define mixed-state form factors of $U(1)$ twist fields. We consider mixed states with density matrices diagonal in the asymptotic particle basis. This includes the thermal Gibbs state as well as all generalized Gibbs ensembles of the Dirac theory. When the mixed state is specialized to a thermal Gibbs state, using a Riemann-Hilbert problem and low-temperature expansion, we obtain finite-temperature form factors of $U(1)$ twist fields. We then propose the expression for form factors of $U(1)$ twist fields in general diagonal mixed states. We verify that these form factors satisfy a system of nonlinear functional differential equations, which is derived from the trace definition of mixed-state form factors. At last, under weak analytic conditions on the eigenvalues of the density matrix, we write down the large distance form factor expansions of two-point correlation functions of these twist fields. Using the relation between the Dirac and Ising models, this provides the large-distance expansion of the R$\acute{\text{e}}$nyi entropy (for integer R$\acute{\text{e}}$nyi parameter) in the Ising model in diagonal mixed states.
\vfill
\end{abstract}

\vspace{1cm} 

\vfill

{\ }\hfill March 2016

\end{titlepage}

\tableofcontents

\section{Introduction}

Correlation functions of local fields play an important role in quantum field theory (QFT) because they yield all physical information of the model and they are directly related to experiment results in condensed matter systems near criticality. However, in general, the evaluation of correlation functions in QFT is a quite non-trivial work. Fortunately, in 1+1 dimensional integrable models of QFT \cite{Karow,Zamo,Smirnov,Delfino,Mussardo}, the existence of an infinite number of conserved charges makes it possible to exactly evaluate many quantities such as form factors, which are matrix elements of local fields in eigenstates of the Hamiltonian.  Successes for these models have been achieved over the last two decades on the determination of vacuum correlation functions via form factor bootstrap approach \cite{Karow,Zamo,Smirnov,Delfino,Mussardo}. For instance, large distance expansions of two-point correlation functions, which are hardly accessible by perturbation theory, can be obtained by form factor expansion (Kallen-Lehmann expansions) \cite{Karow,Zamo,Smirnov,Delfino,Mussardo}under the factorized scattering theory and both their large-distance and short-distance asymptotic behaviors agree with general QFT expectations (see for instance, \cite{ZamoLee,Yurov}).

In recent years, correlation functions in general mixed states have attracted growing interest and triggered an enormous amount of work, because of their wide scope of applications both of theoretical and experimental interest. For instance,  correlation functions in thermal Gibbs state, which can be related to correlation functions on an infinite cylindrical geometry \cite{Matsubara}, have been the subject of intense study in massive integrable QFT (see for instance the review \cite{Ess}).  In particular, the Ising model at finite temperature has been widely investigated by employing several approaches including form factor expansions \cite{Leclair,Fonseca2,Ben1,Mussardo2,Reyes06,Ben2}, integrable differential equations \cite{Fonseca1,Gamsa}, semi-classic methods \cite{Sachdev1}, and the finite volume regularization method \cite{Pozsgay1,Pozsgay2,EssKon1,EssKon2,Pozsgay3,Sz}. On the other hand, more mixed states have been explored, including generalized Gibbs ensembles (GGEs) which have been predicted to occur after quantum quench in integrable models \cite{Rigol,Rigol2,Pol,Fagotti,Fagotti1}, non-equilibrium steady state \cite{Aschbacher2,Ben3,BDaihp,Ben4,DeLuca} and others.  However, further development still needs to be made in order to clarify how the structure of correlation functions depend on the mixed states in general situations. In this paper, we concentrate on correlation functions of  $U(1)$  twist fields in free massive Dirac theory in general mixed states with diagonal density matrix. These density matrices include thermal Gibbs states and generalized Gibbs ensembles.

Twist fields are local fields equipped with non-trivial exchange relations with respect to the fundamental boson fields or fermion fields, associated with a symmetry of the model. The concept of twist fields was first introduced in \cite{Schroer}, as $\mathbb{Z}_2$ monodromy field of the Majorana fermion, corresponding to the spin field of the Ising model.  In the massive Dirac model, there exist a family of bosonic primary twist fields and two families of fermionic primary twit fields, which are associated with the $U(1)$ symmetry transformation. These $U(1)$ twist fields are interacting fields, due to their semi-locality with respect to the Dirac fermions, even though the model we are considering in this paper is a free massive model with a trivial scattering matrix. Vacuum form factors of these fields are well known \cite{Efthimiou,Karowski,Marino} and it has been clear that vacuum correlation functions of $U(1)$ twist fields can be parameterized in terms of solutions to non-linear differential equations \cite{Bernard,Sato,James1,James2}. 

However, there is relatively little known about  correlation functions of $U(1)$ twist fields in general mixed states even though we expect same non-linear differential equations for them. In this paper, we obtain for the first time large distance expansions of two-point correlation functions of $U(1)$ twist fields in general diagonal mixed states, using the method of the ``Liouville space". This method was initially established in \cite{Ben1,Ben2} to derive finite-temperature spin-spin correlation functions,  and then further developed in \cite{Yixiong} to obtain general diagonal mixed-state spin-spin correlation functions, both in the Ising model of QFT. The Liouville space construction \cite{Gelfand,Segal} is based on the GNS construction of $C^*$-algebras \cite{Arveson} and it has applications in thermal and non-equilibrium physics. In the present paper, we apply this method to the free Dirac theory. We define and evaluate the associated mixed-state form factors of $U(1)$ twist fields,  and then formulate mixed-state two-point functions of these fields using form factor expansion with respect to the vacuum in the Liouville space.
 
At zero temperature, the ordinary form factors are mostly obtained by solving a Riemann-Hilbert problem in terms of form factors as functions of rapidities. However, when it comes to the case of general mixed states, this method exhibits some difficulties due to the inability to know the analytic structure of the general density matrix. In this paper, we first evaluate finite-temperature form factors of $U(1)$ twist fields by deriving low-temperature expansions of thermal form factors and setting up a similar Riemann-Hilbert problem as the one in \cite{Ben1,Ben2} for finite-temperature factors in the Ising model. Considering the way how these finite-temperature form factors rely on the eigenvalues of density matrix, we then conjecture an explicit expression for general diagonal mixed-state form factors of $U(1)$ twist fields. These conjectured form factors are proven to be the solution of a system of non-linear first-order functional differential equations for mixed-state form factors. In the derivation of  finite-temperature form factors, we employ imaginary-time formalism \cite{Matsubara} to relate to quantization on the circle and to derive Kubo-Martin-Schwinger (KMS) identity leading to the Riemann-Hilbert problem. But, in the case of general mixed states, we perform real-time manipulation since there is no clear quantization scheme related in imaginary time.

This paper is organized as follows. In section 2, we review the basic concepts of the free massive Dirac theory, and provide a summary of $U(1)$ twist fields and their vacuum form factors. In section 3, we construct the Liouville space and define the mixed-state form factors in the Dirac model. In section 4, we present calculations leading to our main results which include form factors of $U(1)$ twist fields in general diagonal mixed states and the corresponding two-point correlation functions. In section 5, we apply our result of the mixed-state two-point correlation function of $U(1)$ twist fields to compute the R$\acute{\text{e}}$nyi entropy for integer $n$ in the Ising model. Finally, we  conclude in Section 6. 

\section{Dirac theory at zero temperature}
\subsection{Dirac fermions}

In the free massive Dirac theory, fermion operators with evolution in real time $t$ are given in terms of mode operators $D_{\pm}(\theta)$ and $D^\dag_{\pm}(\theta)$:
\beqa \label{fermion}
\Psi_R(x,t)&=&\sqrt m \int d\theta e^{\theta/2}\left(D_+^\dag(\theta)e^{it E_\theta -ixp_\theta}-iD_-(\theta)e^{-it E_\theta +ixp_\theta}\right) \no\\
\Psi_L(x,t)&=&\sqrt m \int d\theta e^{-\theta/2}\left(iD_+^\dag(\theta)e^{it E_\theta -ixp_\theta}-D_-(\theta)e^{-it E_\theta +ixp_\theta}\right)
\eeqa
where 
\beqa
E_\theta&=&m\cosh\theta\;,\no\\
p_\theta&=&m\sinh\theta\no \;.
\eeqa
The Hermitian Conjugate of fermion operators can be obtained by directly taking Hermitian Conjugate of (\ref{fermion}):
\beqa  \label{fermion conjugation}
\Psi^\dag_R(x,t)&=&\sqrt m \int d\theta e^{\theta/2}\left(iD_-^\dag(\theta)e^{it E_\theta -ixp_\theta}+D_+(\theta)e^{-it E_\theta +ixp_\theta}\right) \no\\
\Psi^\dag_L(x,t)&=&\sqrt m \int d\theta e^{-\theta/2}\left(-D_-^\dag(\theta)e^{it E_\theta -ixp_\theta}-iD_+(\theta)e^{-itE_\theta +ixp_\theta}\right)\;.
\eeqa
The creation and annihilation operators satisfy canonical anti-commutation relations:
\beqa \label{algebra}
\{D^\dag_+(\theta_1), D_+(\theta_2)\}&=&\delta(\theta_1-\theta_2)\no\\
\{D^\dag_-(\theta_1), D_-(\theta_2)\}&=&\delta(\theta_1-\theta_2)
\eeqa
with other anti-commutators vanishing.
The fermion operators satisfy the equations of motion
\beqa
\bar\p \Psi_R&=&m\Psi_L,\quad \p \Psi_L=m\Psi_R \no\\
\bar\p \Psi^\dag_R&=&m\Psi^\dag_L,\quad \p \Psi^\dag_L=m\Psi^\dag_R
\eeqa
where we define notations $ \p:=\p_x-\p_t$ and $\b\p:=\p_x+\p_t$, and their anti-commutation relations are
\beqa
\{\Psi_R(x_1),\Psi^\dag_R(x_2)\}&=&4\pi\delta(x_1-x_2)\no\\
\{\Psi_L(x_1),\Psi^\dag_L(x_2)\}&=&4\pi\delta(x_1-x_2)
\eeqa
with other anti-commutators vanishing.
The Hilbert space $\cal H$ is simply the Fock space over algebra (\ref{algebra}) with vacuum state defined by $D_{\pm}|\vac\ket=0$ and with multi-particle states denoted by 
\beq
|\theta_1,\ldots,\theta_N\ket_{\nu_1,\ldots,\nu_N}:=D_{\nu_1}^\dag\cdots D^\dag_{\nu_N}|\vac\ket, \quad\quad \theta_1>\cdots>\theta_N
\eeq
where $\nu_i$ are signs$(\pm)$, corresponding with particle type. Multi-particle states with different ordering can be obtained upon exchange of two particles $(\theta_i,\nu_i)$ and $(\theta_j,\nu_j)$, in agreement with \eqref{algebra}.
The inner products are normalized as follows:
\beq
{}_{\nu_1,\ldots,\nu_N}\bra \theta_1,\ldots,\theta_N|\theta'_1,\ldots,\theta'_N\ket_{\nu'_1,\ldots,\nu'_N}=\prod_{i=1}^N \delta_{\nu_i, \nu'_i}\delta(\theta_i-\theta'_i).
\eeq
Then the resolution of the identity are written as:
\beq
{\bf 1}=\sum_{N=0}^\infty \frac{1}{N!}\sum_{\nu_1,\ldots, \nu_N}\int_{-\infty}^\infty d\theta_1 \cdots \int_{-\infty}^\infty d\theta_N\, |\theta_1,\ldots,\theta_N\ket_{\nu_1,\ldots,\nu_N}\; {}_{\nu_1,\ldots,\nu_N}\bra \theta_1,\ldots,\theta_N|
\eeq
where $N!$ in the denominator comes from overcounting the same state with different orderings of rapidities.

\subsection{Bosonic primary twist fields and their form factors}

The Dirac theory possesses a $U(1)$ internal symmetry $\Psi_{R,L}\mapsto e^{2\pi i \alpha}\Psi_{R,L}$ where $0\leq\alpha<1$ and there exists a family of primary twist fields $\sigma_{\alpha}(x,t)$ associated with this symmetry, which are local, Lorentz spinless, and $U(1)$ neutral, with dimension $\alpha^2$ \cite{Sato}. These fields generate even number of fermions and hence are of bosonic statistics. The bosonic primary twist fields with negative index can be defined by  Hermitian conjugation:
\beq \label{sigma HC}
\sigma^\dag_\alpha=\sigma_{-\alpha}\;,\quad\quad 0\leq\alpha<1\,.
\eeq
Twist fields $\sigma_\alpha$ are associated with branch cuts through which other fields are affected by the $U(1)$ symmetry transformation. These branch cuts,  in principle, can be taken arbitrarily. Here, for convention, we denote by $\sigma^+_\alpha$ the twist fields with branch cuts running towards the right direction, while by $\sigma^-_\alpha$ the ones with branch cuts running towards the left direction.
These two types of twist fields are related to each other by an unitary operator
\beq
Z:=\exp\lt[ \sum_\nu 2\pi i \nu \alpha \int d\theta D^\dag_\nu(\theta)D_\nu(\theta)\rt]
\eeq
which implements the $U(1)$ symmetry transformation and we have
\beq
\sigma^-_\alpha(x,t)=\sigma^+_\alpha (x,t)Z\,.
\eeq
 Twist fields $\sigma^\eta_\alpha$ with $\eta=\pm$ are semi-local with respect to the Dirac fermion fields, and are characterized by equal-time exchange relations
\beq \label{order exchange1}
\renewcommand{\arraystretch}{1.35}
       \Psi_{R,L}(x)\sigma^\eta_\alpha(0) = \Bigg\{
\ba{ll} \lt(\delta_{\eta,-}e^{2\pi i \eta\alpha} +  \delta_{\eta,+}\rt) \sigma^\eta_\alpha(0)\Psi_{R,L}(x) & \quad(x<0) \\
\lt(\delta_{\eta,+}e^{2\pi i \eta \alpha} +  \delta_{\eta,-}\rt) \sigma^\eta_\alpha(0)\Psi_{R,L}(x)  & \quad(x>0)\ea
\eeq
and 
\beq \label{order exchange2}
\renewcommand{\arraystretch}{1.37}
      \Psi^\dag_{R,L}(x)\sigma^\eta_\alpha(0) = \Bigg\{
\ba{ll} \lt(\delta_{\eta,-}e^{-2\pi i \eta\alpha} +  \delta_{\eta,+}\rt) \sigma^\eta_\alpha(0)\Psi^\dag_{R,L}(x) & \quad(x<0) \\
\lt(\delta_{\eta,+}e^{-2\pi i \eta\alpha} +  \delta_{\eta,-}\rt) \sigma^\eta_\alpha(0)\Psi^\dag_{R,L}(x)  & \quad(x>0)\;.\ea
\eeq
Thanks to these twist conditions, two-particle form factors of twist fields $\sigma^{\eta}_\alpha$ for $-1<\alpha<1$ can be fixed, up to normalization, \cite{Schroer,Efthimiou,Karowski,Marino}(see also appendix A of \cite{Bernard}):
\beq
\bra \vac| \sigma^{\eta}_\alpha (0) |\theta_1,\theta_2\ket_{\nu_1,\nu_2}=\delta_{\nu_1,-\nu_2}\ \nu_1\frac{\sin(\pi \alpha)}{2\pi i}\frac{e^{\nu_1\alpha(\theta_1-\theta_2)}}{\cosh\frac{\theta_1-\theta_2}{2}}\,\bra \sigma_\alpha \ket
\eeq
where $\bra \sigma_\alpha \ket:=\bra\vac| \sigma^\eta_\alpha |\vac\ket=c_\alpha m^{\alpha^2}$ is the vacuum expectation value. The dimensionless constants $c_\alpha$ are computed in \cite{Lukyanov,Ben5}. All other higher-particle form factors can be obtained by Wick's theorem due to the fact that twist fields $\sigma^{\eta}_\alpha$ can be expressed as normal-ordered exponentials of bilinear expressions in Dirac fermion operators.  Other matrix elements can be evaluated using crossing symmetry. Note that twist fields $\sigma^{\eta}_\alpha$  have non-zero  form factors only for even particle numbers since they are $U(1)$ neutral. Finally, it is natural for us to define twist fields $\sigma_\alpha$ for all $\alpha\in \mathbb{R}\setminus \mathbb{Z^*}$ by noticing that their form factors  for fixed rapidities are analytic functions of $\alpha$ on $\alpha\in \mathbb{C}\setminus \mathbb{Z^*}$, with general poles on $\mathbb{Z^*}:=\mathbb{Z}\setminus \{0\}$.

\subsection{Fermionic primary twist fields and their form factors}
In the $U(1)$ Dirac theory, there also exist two families of primary twist fields of fermionic statistics, which can be obtained as the coefficients occuring in the operator product expansions (OPEs) of primary twist fields $\sigma_\alpha$ with the Dirac fields $\Psi_R$ and $\Psi^\dag_R$, for all $\alpha\in \mathbb{R}\setminus \mathbb{Z^*}$:
\beqa 
\sigma_{\alpha+1,\alpha}(x,t)&=&\lim_{z\to w}(z-w)^{\alpha} \Psi^\dag_R(x',t')\sigma_\alpha(x,t) \label{mu+1}\label{disorder1}\\
\sigma_{\alpha-1,\alpha}(x,t)&=&\lim_{z\to w}(z-w)^{-\alpha} \Psi_R(x',t')\sigma_\alpha(x,t) \label{mu-1}\label{disorder2}
\eeqa
where $z=-\frac{1}{2}(x'-t')$ and $w=-\frac{1}{2}(x-t)$ with the time ordering $t'>t$. The factors $(z-w)^{\alpha}$ and $(z-w)^{-\alpha}$ are taken on the principal branch. Twist fields $\sigma_{\alpha\pm1,\alpha}$ have charges $\mp1$ , spins $\pm\alpha+1/2$, and dimensions $\alpha^2\pm\alpha+1/2$. Their Hermitian conjugations are given by
\beq \label{mu HC}
\sigma^\dag_{\alpha\pm1,\alpha}=\sigma_{-\alpha\mp1,-\alpha}\,.
\eeq
Again, we define two types of fermionic primary twist fields: $\sigma^+_{\alpha\pm1,\alpha}$ with branch cuts on the right and $\sigma^-_{\alpha\pm1,\alpha}$ with branch cuts on the left, which are related to each other by the unitary operator $Z$
\beq
\sigma^-_{\alpha\pm1,\alpha}=\sigma^+_{\alpha\pm1,\alpha}Z\,.
\eeq
From the definitions (\ref{disorder1}) (\ref{disorder2}) and twist properties of $\sigma_\alpha^\eta$ (\ref{order exchange1}) (\ref{order exchange2}), these twist fields should obey non-trivial equal-time exchange relations with
the Dirac fermion fields
\beq \label{mu twist1}
\renewcommand{\arraystretch}{1.37}
       \Psi_{R,L}(x)\sigma^\eta_{\alpha\pm1,\alpha}(0) = \Bigg\{
\ba{ll} -\lt(\delta_{\eta,-}e^{2\pi i \eta\alpha} +  \delta_{\eta,+}\rt) \sigma^\eta_{\alpha\pm1,\alpha}(0)\Psi_{R,L}(x) & \quad(x<0) \\
-\lt(\delta_{\eta,+}e^{2\pi i \eta\alpha} +  \delta_{\eta,-}\rt) \sigma^\eta_{\alpha\pm1,\alpha}(0)\Psi_{R,L}(x)  & \quad(x>0)\ea
\eeq
and 
\beq  \label{mu twist2}
\renewcommand{\arraystretch}{1.37}
      \Psi^\dag_{R,L}(x)\sigma^\eta_{\alpha\pm1,\alpha}(0) = \Bigg\{
\ba{ll} -\lt(\delta_{\eta,-}e^{-2\pi i \eta\alpha} +  \delta_{\eta,+}\rt) \sigma^\eta_{\alpha\pm1,\alpha}(0)\Psi^\dag_{R,L}(x) & \quad(x<0) \\
-\lt(\delta_{\eta,+}e^{-2\pi i \eta\alpha} +  \delta_{\eta,-}\rt) \sigma^\eta_{\alpha\pm1,\alpha}(0)\Psi^\dag_{R,L}(x)  & \quad(x>0)\;.\ea
\eeq
One-particle form factors of these twist fields can be deduced in the OPEs \cite{James1}:
\beqa
\bra \vac| \sigma^\eta_{\alpha+1,\alpha}(0)|\theta\ket_\nu&=&\delta_{\nu,+}\frac{e^{-i\pi\alpha/2}e^{2\pi i\nu\alpha\delta_{\eta,-}}}{\Gamma(1+\alpha)}m^{\alpha+1/2}e^{(\alpha+1/2)\theta}\bra\sigma_\alpha\ket\\
\bra \vac| \sigma^\eta_{\alpha-1,\alpha}(0)|\theta\ket_\nu&=&-i\delta_{\nu,-}\frac{e^{i\pi\alpha/2}e^{2\pi i\nu\alpha\delta_{\eta,-}}}{\Gamma(1-\alpha)}m^{-\alpha+1/2}e^{(-\alpha+1/2)\theta}\bra\sigma_\alpha\ket\,.
\eeqa
Any higher-particle form factors can be factorised into a product of the associated one-particle form factor and two-particle form factors due to Wick's theorem. Other matrix elements can be obtained by crossing symmetry.

It is worth noting that we can obtain the same families of fermionic primary twist fields $\sigma_{\alpha,\alpha-1}$ and $\sigma_{\alpha,\alpha+1}$ \cite{James1,James2} by shifting $\alpha \mapsto \alpha-1$ in $\sigma_{\alpha+1,\alpha}$ and shifting $\alpha \mapsto \alpha+1$ in $\sigma_{\alpha-1,\alpha}$ respectively. These fields are just a relabelling of the same fermionic primary twist fields.

\section{Liouville space and mixed-state form factors}

The Liouville space $\liou_{\rho}$ for the $U(1)$ Dirac theory is the space of the operators $\End (\mH)$, with inner product specified by the density matrix $\rho$ \
\beq
	{}^{\rho}\bra A| B\ket^\rho = \frc{\Tr\lt(\rho\,A^\dag B\rt)}{
	\Tr\lt(\rho\rt)}
\eeq
where $| A\ket^\rho$ and $| B\ket^\rho$ are the corresponding Liouville states, with $A, B\in\End (\mH)$ . In the present paper, we confine ourselves to the density matrices $\rho$ that are diagonal on the asymptotic state basis:
\beq \label{untwisted DM}
	\rho = \exp\lt[-\int d\theta \,\sum_\nu W_\nu(\theta) \,D_\nu^\dag(\theta) D_\nu(\theta)\rt]
\eeq
where functions $W_\nu(\theta)$ with $\nu=\pm$ are integrable on the real line and make the density matrices well-defined. We consider two cases: the untwisted and twisted cases. In the untwisted case, we consider the density matrix \eqref{untwisted DM}. 
In the twisted case, we consider the density matrix $\rho^\sharp$ with the presence of  two extra unitary operators $e^{2\pi i \nu \alpha\int d\theta D^\dag_\nu(\theta)D_\nu(\theta)}$ which implement the $U(1)$ symmetry:
\beq \label{twisted DM}
	\rho^\sharp = \exp\lt[-\int d\theta \,\sum_\nu W^\sharp_\nu(\theta) \,D_\nu^\dag(\theta) D_\nu(\theta)\rt]
\eeq
where $W^\sharp_\nu(\theta)=W_\nu(\theta)+2\pi i \nu \alpha$.

The Liouville space is spanned by a set of products of creation and annihilation operators in Hilbert space with some particular normalization:
\beq\label{basis}
	|\vac\ket^\rho \equiv {\bf 1},\quad
	|\theta_1,\ldots, \theta_N\ket_{(\nu_1,\ep_1)\ldots(\nu_N,\ep_N)}^\rho
	\equiv Q_{(\nu_1,\ep_1)\ldots(\nu_N,\ep_N)}^\rho(\theta_1,\ldots,\theta_N)\,D^{\ep_1}_{\nu_1}(\theta_1)\cdots D_{\nu_N}^{\ep_N}(\theta_N),
\eeq
with the ordering $\theta_1>\cdots>\theta_N$, where the normalization factors are chosen as
\beq\label{Q}
	Q_{(\nu_1,\ep_1)\ldots(\nu_N,\ep_N)}^\rho(\theta_1,\ldots, \theta_N)\
	:= \prod_{i=1}^N \lt(1+e^{-\ep_i W_{\nu_i}(\theta_i)}\rt),
\eeq 
and where we denote by $D_\nu^-$ the annihilation operators $D_\nu$ and by $D^+_\nu$ the creation operators $D^\dag_\nu$.
 Here, we refer to a doublet $(\nu,\ep)$ as representing the type of a ``Liouville particle'' of rapidity $\theta$. In this sense, our Liouville space can be interpreted as the space of particles and holes excitations from the Liouville vacuum which consists of a number of different types of particles with density $\rho$. Using the cyclic property of the trace and the canonical anti-commutation relations, the inner product of basis states can be deduced as
\beq\
	{}^{\hspace{2.00cm} \rho}_{(\nu_1,\ep_1)\ldots(\nu_N,\ep_N)}\bra\theta_1,\ldots,\theta_N|\theta_1',\ldots,\theta_N'\ket_{(\nu'_1,\ep'_1)\ldots(\nu'_N,\ep'_N)}^\rho
	= \prod_{i=1}^N \lt[\lt(1+e^{-\ep_i W_{\nu_i}(\theta_i)}\rt)
	\delta_{\nu_i ,\nu'_i}\,\delta_{\ep_i,\ep_i'}\,\delta(\theta_i-\theta_i')\rt]
\eeq
with the ordering $\theta_1>\cdots>\theta_N$ and $\theta'_1>\cdots>\theta'_N$.

Every operator $A\in\End({\cal H})$  can be mapped to an operator $A^{\ell}\in\End({\liou_\rho})$ by a linear Liouville left-action 
\beq\label{lefta}
	A^\ell\in \End(\liou_\rho) \;:\; A^\ell|B\ket^\rho = |A B\ket^\rho.
\eeq
To $D_{\nu}^{\ep}(\theta)$, their associated Liouville operators are
\beq \label{liouville mode operators}
\lt[D_{\nu}^{\ep}(\theta)\rt]^{\ell}=\frac{\lZ_{\nu,\ep}^\dag(\theta)}{1+e^{-\ep W_{\nu}(\theta)}}+\frac{\lZ_{\nu,-\ep}(\theta)}{1+e^{\ep W_{\nu}(\theta)}}
\eeq
where $\lZ_{\nu,\ep}^\dag(\theta)$ and $\lZ_{\nu,\ep}(\theta)$ are both defined as Liouville mode operators satisfying anti-commutation relations
\beqa
\{\lZ_{\nu,\ep}(\theta),\lZ_{\nu',\ep'}^\dag(\theta')\}& =& \lt(1+e^{-\ep W_{\nu}(\theta)}\rt)\delta_{\nu, \nu'}\,\delta_{\ep,\ep'}\,\delta(\theta-\theta') \label{anti1}\\
\{\lZ_{\nu,\ep}(\theta),\lZ_{\nu',\ep'}(\theta')\}&=&\{\lZ^\dag_{\nu,\ep}(\theta),\lZ_{\nu',\ep'}^\dag(\theta')\}=0 \label{anti2}\;.
\eeqa
The Liouville space can be seen as the Fock space over this algebra,
\beq
	\lZ_{\nu,\ep}(\theta)|\vac\ket^\rho =0,\quad
	|\theta_1,\ldots,\theta_N\ket_{(\nu_1,\ep_1)\ldots(\nu_N,\ep_N)}^\rho =
	\lZ_{\nu_1,\ep_1}^\dag(\theta_1)\cdots \lZ_{\nu_N,\ep_N}^\dag(\theta_N)
	|\vac\ket^\rho.
\eeq

With the definitions above, it is obvious to see that the mixed-state averages of operators on $\cal H$ are vacuum expectation values on $\liou_\rho$:
\beq \label{HL}
	\bra A\ket_\rho = {}^\rho\bra\vac|A^\ell |\vac\ket^\rho.
\eeq
Using the resolution of the identity in the Liouville space
\beqa \label{resolution}
	{\bf 1}^\ell &=& \sum_{N=0}^\infty \sum_{\nu_1,\ldots,\nu_N} \sum_{\ep_1,\ldots,\ep_N}
	\int_{-\infty}^\infty \lt[ \frc{d\theta_1\cdots d\theta_N}{N!\,
	{\prod_{j=1}^{N}\lt(1+e^{-\ep_j W_{\nu_i}(\theta_j)}\rt)}}\rt.\no \\
	&&\lt.\times |\theta_1,\ldots,\theta_N\ket^\rho_{(\nu_1,\ep_1)\ldots(\nu_N,\ep_N)}
	\;{}_{(\nu_1,\ep_1)\ldots(\nu_N,\ep_N)}^{\hspace{20mm} \rho}\bra\theta_1
	,\ldots,\theta_N|\,\rt]
\eeqa
where $N!$ in the denominator comes from overcounting the same basis state, two-point functions, such as
\[\bra \Or(x,\tau) \Or^\dag(0) \ket_\rho={}^\rho\bra\vac|\Or(x,\tau)^\ell\Or^\dag(0,0)^\ell|\vac\ket^\rho\]
 should have a spectral decomposition on $\liou_\rho$, where we define the matrix elements of left-action operators in the Liouville space as  mixed-state form factors
\beq\label{ff}
	f^{\rho;\Or}_{(\nu_1,\ep_1)\ldots(\nu_N,\ep_N)}(\theta_1,\ldots,\theta_N)
	:= {}^\rho
	\bra\vac|\Or(0,0)^\ell|\theta_1,\ldots,\theta_N\ket_{(\nu_1,\ep_1)\ldots(\nu_N,\ep_N)}^\rho\;.
\eeq
With the definition (\ref{ff}) and anti-commutation relations (\ref{anti2}), the mixed-state form factors satisfy the relations
\beqa \label{exchange theta}
&&f^{\rho;\Or}_{(\nu_1,\ep_1)\cdots(\nu_j,\ep_j)(\nu_{j+1},\ep_{j+1})\cdots(\nu_N,\ep_N)}(\theta_1,\ldots,\theta_j,\theta_{j+1},\ldots,\theta_N)\no \\
&=&-f^{\rho;\Or}_{(\nu_1,\ep_1)\cdots(\nu_{j+1},\ep_{j+1})(\nu_{j},\ep_{j})\cdots(\nu_N,\ep_N)}(\theta_1,\ldots,\theta_{j+1},\theta_{j},\ldots,\theta_N)\;.
\eeqa
The cyclicity of traces leads to the relation
\beq\label{dirac otherside}
	{}^{\hspace{2.1cm} \rho}_{(\nu_1,\ep_1)\cdots(\nu_N,\ep_N)}\bra\theta_1,\ldots,\theta_N|\Or^\ell|\vac\ket^\rho =
	f^{\rho;\Or}_{(\nu_N,-\ep_N)\cdots(\nu_1,-\ep_1)}(\theta_N,\ldots,\theta_1).
\eeq
According to (\ref{HL}), the mixed-state form factors are essentially traces with insertion of operators $D_\nu^\ep(\theta)$, up to an overall factor $Q^\rho_{(\nu_1,\ep_1)\cdots(\nu_N,\ep_N)}(\theta_1,\ldots,\theta_N)$ and up to the subtraction of contact terms at colliding rapidities:
\beqa \label{trace def dirac}
	f^{\rho;\Or}_{(\nu_1,\ep_1)\cdots(\nu_N,\ep_N)}(\theta_1,\ldots,\theta_N)\hspace{6cm}\n
	= \Big[Q^\rho_{(\nu_1,\ep_1)\cdots(\nu_N,\ep_N)}(\theta_1,\ldots,\theta_N)\bra
	\Or\, D_{\nu_1}^{\ep_1}(\theta_1)\cdots D_{\nu_N}^{\ep_N}(\theta_N)\ket_\rho\Big]_{\rm connected}.
\eeqa
 For example, two-particle mixed-state form factors can be written as
\beq
f^{\rho;\Or}_{(\nu_1,\ep_1)(\nu_2,\ep_2)}(\theta_1,\theta_2)=Q^\rho_{(\nu_1,\ep_1)(\nu_2,\ep_2)}(\theta_1,\theta_2)\bra \Or D^{\ep_1}_{\nu_1}(\theta_1)D^{\ep_2}_{\nu_2}(\theta_2)\ket_\rho- {}_{\nu_2,-\ep_2}^{\hspace{6.5mm}\rho}\bra\theta_2|\theta_1\ket^\rho_{\nu_1,\ep_1}\,\bra \Or \ket_\rho\,.
\eeq
Again, using cyclicity of the trace, we  have
\beqa
	{}_{\nu_2,\ep_2}^{\hspace{0.5cm}\rho}\bra\theta_2|\Or^\ell|\theta_1\ket_{\nu_1,\ep_1}^\rho
	&=&
	f_{(\nu_1,\ep_1)(\nu_2,-\ep_2)}^{\rho;\Or}(\theta_1,\theta_2)+
	\bra\Or\ket_\rho\;
	{}_{\nu_2,\ep_2}^{\hspace{5mm}\rho}\bra\theta_2|\theta_1\ket_{\nu_1,\ep_1}^\rho,\label{dirac both side}\\
	{}_{(\nu_2,\ep_2)(\nu_1,\ep_1)}^{\hspace{1.6cm}\rho}\bra\theta_2,\theta_1|\Or^\ell|\vac\ket^\rho
	&=&
	f_{(\nu_1,-\ep_1)(\nu_2,-\ep_2)}^{\rho;\Or}(\theta_1,\theta_2)+
	\bra\Or\ket_\rho\;
	{}_{\nu_2,\ep_2}^{\hspace{5mm}\rho}\bra\theta_2|\theta_1\ket_{\nu_1,-\ep_1}^\rho.\label{dirac left side}
\eeqa
Similar equations for higher numbers of particles can be obtained in the same fashion.

\section{Calculations and main results}
Before we start, let us define the normalized mixed-state form factors
\beqa 
 f^{\eta}_{\nu,\ep}(\theta)&:=&\bra\sigma^\eta_\alpha\ket^{-1}_\rho f^{\rho;  \mu^\eta}_{\nu, \ep }(\theta)\,,\\
f^\eta_{(\nu_1, \ep_1)(\nu_2 ,\ep_2)}(\theta_1,\theta_2)&:=&\bra\sigma^\eta_{\alpha}\ket_\rho^{-1}f^{\rho;\sigma^\eta_{\alpha}}_{(\nu_1, \ep_1)(\nu_2, \ep_2)}(\theta_1, \theta_2) \label{nf2}\,,
\eeqa
where $f^{\rho;  \mu^\eta}_{\nu, \ep }(\theta):= \delta_{\nu,-\ep}f^{\rho;  \sigma^\eta_{\alpha-1,\alpha}}_{\nu ,\ep }(\theta)+\delta_{\nu,\ep}f^{\rho;\sigma^\eta_{\alpha+1,\alpha}}_{\nu, \ep }(\theta) \label{nf1}$
and $\bra\sigma^\eta_\alpha\ket_\rho$ is the normalization,
and their pure-state limits
\beqa
f^{(0)\eta}_{\nu ,\ep }(\theta)&:=&\lim_{W_\pm \to \infty}f^{\eta}_{\nu ,\ep }(\theta)\,,\\
f^{(0)\eta}_{(\nu_1, \ep_1)(\nu_2 ,\ep_2)}(\theta_1, \theta_2)&:=&\lim_{W_\pm \to \infty}f^\eta_{(\nu_1, \ep_1)(\nu_2, \ep_2)}(\theta_1,\theta_2)\,.
\eeqa
Using the trace definition of mixed-state form factors and the ordinary form factors of $U (1)$ twist fields in Hilbert space, we have
\beqa \label{pnf1}
f^{(0)+}_{\nu,\ep}(\theta)&=&\left(-i\ep\delta_{ \nu,-}+\delta_{\nu,+}\right)\frac{e^{- i\pi \nu\alpha/2}}{\Gamma(1+\nu\ep\alpha)}m^{\nu \ep \alpha+1/2}e^{(\nu\ep\alpha+1/2)\theta}\,,\n
 f^{(0)-}_{\nu,\ep}(\theta)&=&f^{(0)+}_{\nu,\ep}(\theta)e^{2\pi i\nu\alpha\delta_{\ep,+}}
\eeqa
and
\beqa \label{pnf2}
f^{(0)\eta}_{(\nu_1 ,\ep_1)(\nu_2 ,\ep_2)}(\theta_1, \theta_2)&=&\delta_{\ep_1, \ep_2}\delta_{\nu_1 ,-\nu_2}\,\nu_1 \ep_1 \frac{\sin(\pi\alpha)}{2\pi i}\frac{e^{\nu_1 \ep_1\alpha(\theta_1-\theta_2)}}{\cosh(\frac{\theta_1-\theta_2}{2})}+\no\\
&&\delta_{\ep_1 ,-\ep_2}\delta_{\nu_1, \nu_2}\,i\nu_1  \frac{\sin(\pi\alpha)}{2\pi i} \frac{e^{\nu_1 \ep_1\alpha(\theta_1-\theta_2)}e^{-i\pi\eta\nu_1 \alpha}}{\sinh\lt(\frac{\theta_1-\theta_2+\eta i(\ep_1-\ep_2)0^+}{2}\rt)}\,.
\eeqa

\subsection{Finite-temperature form factors of $U(1)$ twist fields }

\subsubsection{Riemann-Hilbert problem}

Since our mixed-state form factors are defined as traces, one traditional way to evaluate them is just directly performing the traces with the help of  the known matrix elements on Hilbert space. However, this procedure  involves a re-summation of disconnected terms, leaving a quite complicated computation. Particularly, for $U(1)$ twist fields of current interest, the re-summation will be infinite, because twist fields are normal-ordered exponentials of bilinear combinations and they have non-zero matrix elements for arbitrarily large number of particles. Therefore, this method seems to break down for evaluating  mixed-state form factors of $U(1)$ twist fields.

At zero temperature, form factors are functions of rapidities and they satisfy a set of analytic conditions which form the Riemann-Hilbert problem. By solving these equations, form factors can be fully fixed up to normalizations. Unfortunately, even in our Louville space set-up, this technique is hardly applicable to the evaluation of mixed-state form factors because the analytic structure of the function $W_\nu(\theta)$ in general is not accessible. However, we can begin by calculating finite-temperature form factors of $U(1)$ twist fields and a  Riemann-Hilbert problem can be derived in terms of thermal one-particle form factors.

In thermal Gibbs state with the untwisted density matrix, where $W_\pm(\theta)=LE_\theta$ ($L$ is inverse temperature), we consider two-point functions in imaginary-time formalism 
\beq
g(x,\tau)=\;^{\rho}\bra \vac| \sigma^\eta_{\lt(\alpha-\nu(-1),\alpha\rt)}\,(0,0)^{\ell} \,\Psi^{-\nu}_R (x,\tau)^{\ell}|\vac \ket^{\rho}
\eeq
where we denote by $\Psi^\mp$ the fermion operator $\Psi$ and its Hermitian Conjugation $\Psi^\dag$ respectively.
Using Dirac fermions' mode expansions (\ref{fermion}),  (\ref{fermion conjugation}), and Liouville left-action , two-point functions $g(x,\tau)$ can be written as  finite-temperature form factor expansions:
\beq
g(x,\tau)=\sqrt m \int d\theta e^{\theta/2}\left[i\frac{f^{\rho;  \mu^\eta}_{\nu, + }(\theta)}{1+e^{-LE_\theta}}e^{\tau E_\theta -ixP_\theta}+\frac{f^{\rho;  \mu^\eta}_{-\nu ,- }(\theta)}{1+e^{LE_\theta}}e^{-\tau E_\theta +ixP_\theta}\right] \,.
\eeq
Thanks to twist conditions (\ref{mu twist1}) and (\ref{mu twist2}), we can  derive KMS relations at finite temperature in the Dirac theory:
\beqa
g(x,\tau)&=&-\lt(\delta_{\eta,+}\,e^{-\eta\nu 2\pi i \alpha}+\delta_{\eta,-}\rt)g(x,\tau-L)\qquad (x>0) \label{KMS1}\\
g(x,\tau)&=&-\lt(\delta_{\eta,-}\,e^{-\eta\nu 2\pi i \alpha}+\delta_{\eta,+}\rt)g(x,\tau-L) \qquad  (x<0)  \label{KMS2}\,.
\eeqa
To make sense of KMS relations (\ref{KMS1}) and (\ref{KMS2}), one-particle form factor $f^{\rho;  \mu^\eta}_{\nu ,+ }(\theta)$ and $f^{\rho;  \mu^\eta}_{-\nu ,- }(\theta)$ should satisfy the following requirements:
\begin{enumerate}
\item Analytic structure: $f^{\rho;  \mu^\eta}_{\nu, + }(\theta)$ and $f^{\rho;  \mu^\eta}_{-\nu ,- }(\theta)$ are analytic as functions of $\theta$ on the complex plane except at some simple poles. Analytic structure is specialized in the region ${\rm Im(\theta)}\in[-i \pi,i\pi]$:
\begin{enumerate}
\item 
Thermal poles and zeroes: 
\\\\
$f^{\rho;  \mu^\eta}_{\nu, + }(\theta)$ has poles at \\
                   \[\theta=\gamma^{ \nu}_n-\eta\frac{i\pi }{2},\quad n\in \mathbb{Z}+\frac{1}{2}\]

and zeroes at 
\[\theta=\lam_n-\eta\frac{i\pi }{2},\quad n\in \mathbb{Z}+\frac{1}{2}\,;\]

        $f^{\rho;  \mu^\eta}_{-\nu ,- }(\theta)$ has poles at \\
                   \[\theta=\gamma^{ \nu}_n+\eta\frac{i\pi }{2},\quad n\in \mathbb{Z}+\frac{1}{2}\]

and zeroes at 
\[\theta=\lam_n+\eta\frac{i\pi }{2},\quad n\in \mathbb{Z}+\frac{1}{2}\]
where \[ \sinh \gamma^{ \nu}_n=\frac{2\pi(n+\nu\alpha)}{mL}\,,\quad \sinh\lam_n=\frac{2\pi n}{mL}\,,\]

\item 
$f^{\rho;  \mu^\eta}_{\nu ,+ }(\theta)$ and $f^{\rho;  \mu^\eta}_{-\nu ,- }(\theta)$ are related by relations
\beq \label{half pi i}
f^{\rho;  \mu^\eta}_{\nu ,+ }(\theta\pm i\pi/2) =\pm f^{\rho;  \mu^\eta}_{-\nu ,- }(\theta\mp i\pi/2)
\eeq
for all $\theta$ except $\theta=\gamma^{ \nu}_n, \, n\in \mathbb{Z}+\frac{1}{2}$.
\end{enumerate}

\item  Crossing symmetry:
\beqa \label{crossing}
f^{\rho;  \mu^\eta}_{\nu ,+ }(\theta \pm i\pi)=\pm f^{\rho;  \mu^\eta}_{-\nu ,- }(\theta)\label{cs1}\,.
\eeqa
\item Quasi-periodicity:
\beqa \label{qp}
f^{\rho;  \mu^\eta}_{\ep\nu ,\ep }(\theta \pm 2i\pi)=-f^{\rho;  \mu^\eta}_{\ep\nu ,\ep }(\theta) \label{qp1} \,.
\eeqa
\end{enumerate}
The derivation of a similar Riemann-Hilbert problem can be found in \cite{Ben1}.
Taking into account the properties mentioned above as well as the fact that mixed-state form factors reproduce the ordinary form factors in Hilbert space under the limit $W_\pm(\theta)\to \infty$, and by analogy with thermal form factors of twist fields in Ising model in \cite{Ben1,Ben2},
we conjecture that finite-temperature one-particle form factors of the fermionic $U(1)$ twist field are expressed as a product of the so called ``leg-factor" \cite{Ben1,Ben2,Yixiong} and  normalized vacuum one-particle form factor, up to the overall normalization $\bra\sigma^\eta_\alpha \ket_\rho$ :
\beqa 
f^{\rho;  \mu^\eta}_{\nu, \ep }(\theta)=f^{\eta}_{\nu.\ep}(\theta)\bra\sigma^\eta_\alpha \ket_\rho=f^{(0)\eta}_{\nu,\ep}(\theta)h^\eta_{\nu,\ep}(\theta) \bra\sigma^\eta_\alpha \ket_\rho \label{c1}
\eeqa
with $h^\eta_{\nu ,\ep}(\theta)$ the leg-factors: 
\beqa
h^\eta_{\nu, \ep}(\theta)&=&\exp\Bigg[\int \frac{d\theta'}{2\pi i}\frac{A^\eta_{\nu,\ep}(\theta,\theta')}{\cosh(\frac{\theta-\theta'}{2})}\log\left(\frac{1+e^{-LE_{\theta'}}}{1+e^{2\pi i \eta\nu \alpha}e^{-LE_{\theta'}}}   \right)\no\\
&&+\int_{-\infty-\eta\ep i0^+}^{\infty-\eta\ep i0^+}\frac{d\theta'}{2\pi i}\frac{B^\eta_{\nu,\ep}(\theta,\theta')}{\sinh(\frac{\theta-\theta'}{2})}\log\left(\frac{1+e^{-LE_{\theta'}}}{1+e^{-2\pi i \eta\nu\alpha}e^{-LE_{\theta'}}}   \right)\Bigg]\,.                                                                                                                                                                                                                                                                                                                                                                                                                                                                                                                                                                                                                                                                                                                                                                                                                                                                                                                                                                                                                                                                                                                                                                                                                                                                                                                                                                                                                                                                                                                                                                                                                                                                                                                                                                                                                                                                             
\eeqa
Factors $A^\eta_{\nu,\ep}(\theta,\theta')$ and $B^\eta_{\nu,\ep}(\theta,\theta')$, due to \eqref{half pi i},\eqref{crossing} and \eqref{qp},  must satisfy a set of relations
\begin{enumerate}

\item 
\beqa \label{pi/2}
A^\eta_{\nu,\ep}(\theta\pm i\pi /2,\theta\pm i\pi /2)&=&B^\eta_{\nu,\ep}(\theta\pm i\pi /2,\theta\pm i\pi /2)=-\eta\ep/2\n
A^\eta_{\nu,\ep}(\theta\pm i\pi /2,\theta')&=&\pm i B^\eta_{-\nu,-\ep}(\theta\mp i\pi /2,\theta')\n
B^\eta_{\nu,\ep}(\theta\pm i\pi /2,\theta')&=&\pm i A^\eta_{-\nu,-\ep}(\theta\mp i\pi /2,\theta')\,.
\eeqa
\item
\beqa  \label{pi}
A^\eta_{\nu,\ep}(\theta\pm i\pi ,\theta)&=&\pm \eta\ep i/2\n
B^\eta_{\nu,\ep}(\theta\pm i\pi ,\theta\pm i\pi )&=&- \eta\ep/2\n
A^\eta_{\nu,\ep}(\theta\pm i\pi ,\theta')&=&\pm i B^\eta_{-\nu,-\ep}(\theta,\theta')\n
B^\eta_{\nu,\ep}(\theta\pm i\pi ,\theta')&=&\pm i A^\eta_{-\nu,-\ep}(\theta,\theta')\,.
\eeqa

\item
\beqa \label{2pi}
A^\eta_{\nu,\ep}(\theta\pm2 i\pi ,\theta\pm i\pi )&=&\pm \eta\ep i/2 \n
B^\eta_{\nu,\ep}(\theta\pm2 i\pi ,\theta)&=&\eta\ep/2\,.
\eeqa

\end{enumerate}

To fully determine factors  $A^\eta_{\nu,\ep}(\theta,\theta')$ and $B^\eta_{\nu,\ep}(\theta,\theta')$ , it is intuitive to exploit low-temperature expansions of finite-temperature form factors, which will be discussed  in the next subsection.

\subsubsection{Low-temperature expansion}

Using the trace definition of mixed-state form factors and the  factorisation of higher-particle twist field form factors, we can deduce low-temperature expansions of one- and two-particle normalized form factors for $U(1)$ twist fields in powers of $e^{-mL}$ as follows:
\beqa
f^\eta_{\nu, \ep }(\theta)&=&f^{(0)\eta}_{\nu, \ep }(\theta)\no\\
&&+\sum_{\nu'}\int d\theta' e^{-mL \cosh\theta'}\Big[ f^{(0)\eta}_{(\nu ,\ep)(\nu' ,+)}(\theta, \theta')f^{(0)\eta}_{\nu', -}( \theta')-\no\\
&&f^{(0)\eta}_{(\nu, \ep)(\nu', -)}(\theta, \theta')f^{(0)\eta}_{\nu', +}( \theta')\Big]
+O(e^{-2mL})\label{low-t1}
\eeqa
and
\beqa
f^\eta_{(\nu_1 ,\ep_1)(\nu_2, \ep_2)}(\theta_1, \theta_2)&=&f^{(0)\eta}_{(\nu_1, \ep_1)(\nu_2, \ep_2)}(\theta_1 ,\theta_2)\no\\
&&+\sum_{\nu} \int d\theta e^{-mL\cosh\theta}\Big[f^{(0)\eta}_{(\nu_1, \ep_1)(\nu, +)}(\theta_1, \theta)f^{(0)\eta}_{(\nu_2, \ep_2)(\nu, -)}(\theta_2, \theta)\no\\
&&-f^{(0)\eta}_{(\nu_1, \ep_1)(\nu ,-)}(\theta_1, \theta)f^{(0)\eta}_{(\nu_2 ,\ep_2)(\nu, +)}(\theta_2, \theta)\Big]+O(e^{-2mL})\,.
\eeqa
Then, we turn our attention to the expression of $f^\eta_{\nu ,\ep }(\theta)$ in our conjecture (\ref{c1}).
We taylor expand in $h^\eta_{\nu ,\ep}(\theta)$ the two logarithmic terms  as functions of $e^{-mL}$ at the point $e^{-mL}=0$ in the low-temperature limit, and we then have
\beqa
f^\eta_{\nu, \ep }(\theta)&=&f^{(0)\eta}_{\nu ,\ep }(\theta)\no\\
&&+f^{(0)\eta}_{\nu, \ep }(\theta)\Bigg[\int \frac{d\theta'}{2\pi i}\frac{A^\eta_{\nu,\ep}(\theta,\theta')}{\cosh(\frac{\theta-\theta'}{2})}(1-e^{2\pi i \eta\nu\alpha})e^{-LE_{\theta'}}\no\\
&&+\int \frac{d\theta'}{2\pi i}\frac{B^\eta_{\nu,\ep}(\theta,\theta')}{\sinh(\frac{\theta-\theta'}{2})}(1-e^{-2\pi i \eta\nu\alpha})e^{-LE_{\theta'}}\Bigg]\no\\
&&+O(e^{-2mL})\label{low-t2}\,.
\eeqa
By comparing (\ref{low-t1}) and (\ref{low-t2}), we arrive at
\beqa
A^\eta_{\nu,\ep}(\theta,\theta')=B^\eta_{\nu,\ep}(\theta,\theta')=-\eta\ep\,\frac{1}{2}e^{\frac{(\theta'-\theta)}{2}}
\eeqa
which are in agreement with relations \eqref{pi/2}, \eqref{pi} and \eqref{2pi}.
Consequently, finite-temperature one-particle form factors are fully obtained as:
\beq \label{Gibbs1}
f^{\rho;  \mu^\eta}_{\nu, \ep }(\theta)=f^{(0)\eta}_{\nu,\ep}(\theta)h^\eta_{\nu,\ep}(\theta) \bra\sigma^\eta_\alpha \ket_\rho 
\eeq
with
\beqa \label{thermal leg}
h^\eta_{\nu, \ep}(\theta)&=&\exp\Bigg[-\eta\ep\int \frac{d\theta'}{2\pi i}\frac{\frac{1}{2}e^{\frac{(\theta'-\theta)}{2}}}{\cosh(\frac{\theta-\theta'}{2})}\log\left(\frac{1+e^{-LE_{\theta'}}}{1+e^{2\pi i \eta\nu \alpha}e^{-LE_{\theta'}}}   \right)\no\\
&&-\eta\ep \int_{-\infty-\eta\ep i0^+}^{\infty-\eta\ep i0^+}\frac{d\theta'}{2\pi i}\frac{\frac{1}{2}e^{\frac{(\theta'-\theta)}{2}}}{\sinh(\frac{\theta-\theta'}{2})}\log\left(\frac{1+e^{-LE_{\theta'}}}{1+e^{-2\pi i \eta\nu\alpha}e^{-LE_{\theta'}}}   \right)\Bigg]\,.                                                                                                                                                                                                                                                                                                                                                                                                                                                                                                                                                                                                                                                                                                                                                                                                                                                                                                                                                                                                                                                                                                                                                                                                                                                                                                                                                                                                                                                                                                                                                                                                                                                                                                                                                                                                                                                                             
\eeqa
This solution is uniquely fixed, up to a normalization, by the asymptotic behavior  $f^{\rho;  \mu^\eta}_{\nu, \ep }(\theta)\sim O(1)$ at $|\theta|\rightarrow\infty$, since the fermionic primary twist field is a primary field of spin $0$.
Similarly, we postulate that  finite-temperature two-particle form factors of $U(1)$ primary twist fields have the structure:
\beq \label{Gibbs2}
f^{\rho;\sigma^\eta_\alpha}_{(\nu_1, \ep_1)(\nu_2, \ep_2)}(\theta_1,\theta_2)=f^{(0)\eta}_{(\nu_1, \ep_1)(\nu_2 ,\ep_2)}(\theta_1,\theta_2)h^\eta_{\nu_1, \ep_1}(\theta_1)h^\eta_{\nu_2 ,\ep_2}(\theta_2) \bra\sigma^\eta_\alpha \ket_\rho 
\eeq
which indeed reproduce the correct ordinary two-particle form factors in the pure-state limit. 

\subsection{Mixed-state form factors of $U(1)$ twist fields}
\subsubsection{Exact mixed-state form factors of $U(1)$ twist fields}
As we see, in the case of thermal Gibbs state, we can formulate a set of equations and analytic conditions  by setting up a Riemann-Hilbert problem. The minimal solutions are finite-temperature one-particle form factors. In the case of general diagonal mixed-state, as we mentioned before, such techniques can not be employed, because we can not assume any analytic property of general $W_{\nu}(\theta)$. However, after observing that thermal form factors (\ref{Gibbs1}) and (\ref{Gibbs2}) depend on function $W_{\nu}(\theta)$ in a trivial way,  it is reasonable to replace in leg-factors (\ref{thermal leg}) $LE_{\theta}$ with $W_{\nu}(\theta)$ or $W_{-\nu}(\theta)$, leading to the following proposition: 
\begin{propo}\label{p}{\ }
\\\\
\noindent {\bf I.} The diagonal mixed-state one- and two-particle form factors of $U(1)$ twist fields are given by
\beqa
f^{\rho;  \mu^\eta}_{\nu, \ep }(\theta)&=&f^{(0)\eta}_{\nu,\ep}(\theta)h^\eta_{\nu,\ep}(\theta) \bra\sigma^\eta_\alpha \ket_\rho   \label{mixed-state ff1}\\
&&\no\\
f^{\rho;\sigma^\eta_\alpha}_{(\nu_1, \ep_1)(\nu_2, \ep_2)}(\theta_1,\theta_2)&=&f^{(0)\eta}_{(\nu_1 ,\ep_1)(\nu_2, \ep_2)}(\theta_1,\theta_2)h^\eta_{\nu_1, \ep_1}(\theta_1)h^\eta_{\nu_2, \ep_2}(\theta_2) \bra\sigma^\eta_\alpha \ket_\rho \label{mixed-state ff2}
\eeqa
where
\beqa \label{leg}
h^\eta_{\nu, \ep}(\theta)&=&\exp\Bigg[-\eta\ep\int \frac{d\theta'}{2\pi i}\frac{\frac{1}{2}e^{\frac{(\theta'-\theta)}{2}}}{\cosh(\frac{\theta-\theta'}{2})}\log\left(\frac{1+e^{-W_{-\nu}(\theta')}}{1+e^{2\pi i \eta\nu \alpha}e^{-W_{-\nu}(\theta')}}   \right)\no\\
&&-\eta\ep \int_{-\infty-\eta\ep i0^+}^{\infty-\eta\ep i0^+}\frac{d\theta'}{2\pi i}\frac{\frac{1}{2}e^{\frac{(\theta'-\theta)}{2}}}{\sinh(\frac{\theta-\theta'}{2})}\log\left(\frac{1+e^{-W_{\nu}(\theta')}}{1+e^{-2\pi i \eta\nu\alpha}e^{-W_{\nu}(\theta')}}   \right)\Bigg]\,.                                                                                                                                                                                                                                                                                                                                                                                                                                                                                                                                                                                                                                                                                                                                                                                                                                                                                                                                                                                                                                                                                                                                                                                                                                                                                                                                                                                                                                                                                                                                                                                                                                                                                                                                                                                                                                                                             
\eeqa

\noindent {\bf II.}  Higher-particle form factors can be evaluated by using Wick's theorem on the particles. The overall normalization is $\bra\sigma^\eta_\alpha\ket_\rho$, the contraction of two particles $(\theta_1,\nu_1,\ep_1)$ and $(\theta_2,\nu_2,\ep_2)$ is given by the normalized two-particle form factor $f^{\eta}_{(\nu_1, \ep_1)(\nu_2, \ep_2)}(\theta_1,\theta_2)$, and the remaining single particle $(\theta,\nu,\ep)$, if any, gives a factor $f^{\eta}_{\nu,\ep}(\theta)$; further, there is a minus sign for every crossing of contractions.

\noindent {\bf III.} Form factors of $U(1)$ twist fields are analytic functions, except for possible ``kinematic poles'' at colliding rapidities, in the strip  ${\rm Im}(\theta_j)\in(0,\pi)$ for $\eta\ep_j=+$ and ${\rm Im}(\theta_j)\in(-\pi,0)$ for $\eta\ep_j=-$. Leg-factors $h^\eta_{\nu,\ep}(\theta)$, as functions of $\theta\in\R$, are ordinary integrable functions obtained by continuous continuation from these analyticity regions, and they satisfy
\beq
h^\eta_{\nu,\ep}(\theta)h^\eta_{\nu,-\ep}(\theta)=\frac{1+e^{-W_{\nu}(\theta)}}{1+e^{-2\pi i \eta\nu \alpha}e^{-W_{\nu}(\theta)}}\,.
\eeq

\end{propo}
It is a simple matter to check that mixed-state form factors above do agree with \eqref{exchange theta} considering \eqref{pnf1},\eqref{pnf2}, and with \eqref{dirac otherside} using \eqref{sigma HC}, \eqref{mu HC}, and complex conjugation. 

Concerning the normalization $\bra \sigma^\eta_\alpha \ket_\rho$, it has not been exactly calculated so far. In analogy to the computation of $c_\alpha$ in \cite{James1}, we obtain a same recursion relation for the normalization:
\beq \label{rr}
\frac{\bra \sigma^\eta_{\alpha+1} \ket_\rho}{\bra \sigma^\eta_{\alpha} \ket_\rho}=\frac{\Gamma(-\alpha)}{\Gamma(1+\alpha)}m^{2\alpha+1}\,
\eeq
which, in the pure-state limit, is in agreement with the result of \cite{James1} obtaind in the $U(1)$ Dirac model at zero temperature
\[
\frac{\bra \sigma_{\alpha+1} \ket}{\bra \sigma_{\alpha} \ket}=\frac{\Gamma(-\alpha)}{\Gamma(1+\alpha)}m^{2\alpha+1}\,.
\]
However, to fully determine the normalization $\bra \sigma^\eta_\alpha \ket_\rho$, one needs to find the initial condition of this recursive relation which involves the function  $W_\nu(\theta)$ indicating the mixed states. Once the normalization for $\alpha\in[0,1/2]$ is known, it is known for $\alpha\in[-1/2,1/2]$ by conjugation, and then known for all $\alpha$ thanks to this recursion relation. The derivation of \eqref{rr} is presented in appendix \ref{recursion}.

\subsubsection{Non-linear functional differential system of equations}

As we recall, $U(1)$ twist fields are in the  form of normal-ordered exponential of bilinear combinations. For instance, the twist field $\sigma^\eta_\alpha$ is given by
\beq \label{order field}
\sigma_\alpha^\eta=\bra \sigma_\alpha \ket \lt(:\exp\lt[ \sum_{(\nu_1,\ep_1)(\nu_2,\ep_2)}\int d\theta_1d\theta_2 F^\eta_{(\nu_1,\ep_1)(\nu_2,\ep_2)}(\theta_1,\theta_2)D^{\ep_1}_{\nu_1}(\theta_1)D^{\ep_2}_{\nu_2}(\theta_2)\rt]:\rt)
\eeq
with $F^\eta_{(\nu_1,\ep_1)(\nu_2,\ep_2)}(\theta_1,\theta_2)=-\frac{1}{2}f^{(0)\eta}_{(\nu_1,-\ep_1)(\nu_2,-\ep_2)}(\theta_1,\theta_2)$. Thus, proposition \ref{p} can be verified by mimicking arguments employed in \cite{Yixiong}. We deduce, from the trace definition and Wick's theorem, a system of non-linear functional differential equations for $U(1)$ twist fields mixed-state form factors as functions of $W_{\nu}(\theta)$.

\subsubsection*{\bf{Derivation}}

We denote
\beqa
\tilde f^{\eta}_{\nu,\ep}(\theta)&:=&Q^\rho_{\nu,\ep}(\theta)\frac{\Tr \lt(\rho\, \lt(\delta_{\nu,-\ep}\,\sigma^\eta_{\alpha-1,\alpha}+\delta_{\nu,\ep}\,\sigma^\eta_{\alpha+1,\alpha}\rt) D^\ep_\nu(\theta) \rt)}{\Tr\lt(\rho \sigma^\eta_\alpha\rt)}\\
\tilde f^{\eta}_{(\nu_1,\ep_1)(\nu_2,\ep_2)}(\theta_1,\theta_2)&:=&Q^\rho\twopair(\theta_1,\theta_2)\frac{\Tr\lt(\rho \sigma^\eta_\alpha D^{\ep_1}_{\nu_1}(\theta_1)D^{\ep_2}_{\nu_2}(\theta_2)\rt)}{\Tr\lt(\rho \sigma^\eta_\alpha\rt)}
\eeqa
and the notations are similar for higher numbers of insertions of creation and annihilation operators. Using 
\beq
\frac{\p \rho}{\p W_\nu(\theta)}=-\rho D^+_\nu(\theta)D_\nu(\theta)\,,
\eeq
we have
\beqa
&&\hspace{-1.7cm}\frac{\p }{\p W_{\nu'}(\beta)}\frac{\Tr \lt(\rho \lt(\delta_{\nu,-\ep}\,\sigma^\eta_{\alpha-1,\alpha}+\delta_{\nu,\ep}\,\sigma^\eta_{\alpha+1,\alpha}\rt) D^\ep_\nu(\theta) \rt)}{\Tr\lt(\rho \sigma^\eta_\alpha\rt)}\n
&=&-\frac{\Tr\lt(\rho \lt(\delta_{\nu,-\ep}\,\sigma^\eta_{\alpha-1,\alpha}+\delta_{\nu,\ep}\,\sigma^\eta_{\alpha+1,\alpha}\rt) D^\ep_\nu(\theta)D^+_{\nu'}(\beta)D_{\nu'}(\beta)\rt)}{\Tr\lt(\rho \sigma^\eta_\alpha\rt)}\n
&&+\frac{\Tr\lt(\rho \lt(\delta_{\nu,-\ep}\,\sigma^\eta_{\alpha-1,\alpha}+\delta_{\nu,\ep}\,\sigma^\eta_{\alpha+1,\alpha}\rt) D^\ep_\nu(\theta)\rt)}{\Tr\lt(\rho \sigma^\eta_\alpha\rt)}\frac{\Tr\lt(\rho \sigma^\eta_\alpha D^+_{\nu'}(\beta)D_{\nu'}(\beta)\rt)}{\Tr\lt(\rho \sigma^\eta_\alpha\rt)}\n
    &=&\frac{\tilde f^{\eta}_{\nu,\ep}(\theta) \tilde f^{\eta}_{(\nu',+)(\nu',-)}(\beta,\beta)-\tilde f^{\eta}_{(\nu,\ep)(\nu',+)(\nu',-)}(\theta,\beta,\beta)}{Q^\rho_{(\nu,\ep)(\nu',+)(\nu',-)}(\theta,\beta,\beta)}\n
&=&\frac{\tilde f^{\eta}_{\nu',+}(\beta) \tilde f^{\eta}_{(\nu,\ep)(\nu',-)}(\theta,\beta)-\tilde f^{\eta}_{\nu',-}(\beta) \tilde f^{\eta}_{(\nu,\ep)(\nu',+)}(\theta,\beta)}{Q^\rho_{(\nu,\ep)(\nu',+)(\nu',-)}(\theta,\beta,\beta)}
\eeqa
where we use Wick's theorem in the last step. By recalling the definition \eqref{Q}, we find
\beq
\lt(\frac{\p}{\p W_{\nu'}(\beta)}+\frac{\ep \delta_{\nu,\nu'}\delta(\theta-\beta)}{Q^\rho_{\nu,-\ep}(\theta)}\rt)\tilde f^{\eta}_{\nu,\ep}(\theta)=\frac{\tilde f^{\eta}_{\nu',+}(\beta) \tilde f^{\eta}_{(\nu,\ep)(\nu',-)}(\theta,\beta)-\tilde f^{\eta}_{\nu',-}(\beta) \tilde f^{\eta}_{(\nu,\ep)(\nu',+)}(\theta,\beta)}{4\cosh^2\lt(\frac{W_{\nu'}(\beta)}{2}\rt)}\,.
\eeq
We then differentiate \[\frac{\Tr\lt(\rho \sigma^\eta_\alpha D^{\ep_1}_{\nu_1}(\theta_1)D^{\ep_2}_{\nu_2}(\theta_2)\rt)}{\Tr\lt(\rho \sigma^\eta_\alpha\rt)}\] with respect to $W_{\nu'}(\beta)$, and we find, following the same lines,
\beqa
&&\hspace{-2.1cm}\lt(\frac{\p}{\p W_\nu(\beta)}+\frac{\ep_1\delta_{\nu,\nu_1}\delta(\beta-\theta_1)}{1+e^{\ep_1W_{\nu_1}(\theta_1)}}+\frac{\ep_2\delta_{\nu,\nu_2}\delta(\beta-\theta_2)}{1+e^{\ep_2W_{\nu_2}(\theta_2)}}\rt) \tilde f^{\eta}\twopair (\theta_1,\theta_2)\n
\hspace{-1.1cm}&=&\frac{\tilde f^{\eta}_{(\nu_1,\ep_1)(\nu,+)}(\theta_1,\beta) \tilde f^{\eta}_{(\nu_2,\ep_2)(\nu,-)}(\theta_2,\beta)-
\tilde f^{\eta}_{(\nu_1,\ep_1)(\nu,-)}(\theta_1,\beta) \tilde f^{\eta}_{(\nu_2,\ep_2)(\nu,+)}(\theta_2,\beta)}{4\cosh^2\lt(\frac{W_{\nu'}(\beta)}{2}\rt)}\,.
\eeqa
Finally, by using the relations
\beqa
&&\tilde f^{\eta}_{\nu ,\ep}(\theta)=f^{\eta}_{\nu ,\ep}(\theta),\n
&&\tilde f^{\eta}_{(\nu_1,\ep_1)(\nu_2,\ep_2)}(\theta_1,\theta_2)=f^{\eta}_{(\nu_1,\ep_1)(\nu_2,\ep_2)}(\theta_1,\theta_2)+(1+e^{-\ep_1W_{\nu_1}(\theta_1)})\delta_{\nu_1,\nu_2}\delta_{\ep_1,-\ep_2}\delta(\theta_1-\theta_2),\no
\eeqa
we obtain a system of functional differential equations for the mixed-state form factors of $U(1)$ twist fields:
\beqa
\frac{\p f^{\eta}_{\nu ,\ep}(\theta)}{\p W_{\nu'}(\beta)}&=&\frac{f^{\eta}_{\nu',+}(\beta)f^{\eta}_{(\nu, \ep)(\nu',-)}(\theta, \beta)-f^{\eta}_{\nu',-}(\beta)f^{\eta}_{(\nu, \ep)(\nu',+)}(\theta, \beta)}{4\cosh^2\left(\frac{W_{\nu'}(\beta)}{2}\right)}\label{pde1}\\
\frac{\p f^{\eta}_{(\nu_1, \ep_1)(\nu_2 ,\ep_2)}(\theta_1, \theta_2)}{\p W_{\nu}(\beta)}&=&\frac{f^{\eta}_{(\nu_1, \ep_1)(\nu, +)}(\theta_1 ,\beta)f^{\eta}_{(\nu_2 ,\ep_2)(\nu, -)}(\theta_2, \beta)-f^{\eta}_{(\nu_1, \ep_1)(\nu, -)}(\theta_1 ,\beta)f^{\eta}_{(\nu_2, \ep_2)(\nu,+)}(\theta_2, \beta)}{4\cosh^2\left(\frac{W_{\nu}(\beta)}{2}\right)}.\n
&&\label{pde2}
\eeqa 
This system of functional differential equations enjoys the virtue that it does not require the analytic structure of $W_{\nu}(\theta)$ and it works for general $W_{\nu}(\theta)$.

\subsubsection*{\bf{Uniqueness}}

We can deduce, from the trace definition of mixed-state form factors, the large-$W_\nu(\theta)$ expansions of one-and two-particle mixed-state form factors
\beqa\label{LW1}
f^\eta_{\nu,\ep}(\theta)&=&f_{\nu,\ep}^{(0)\eta}(\theta) + \sum_{\nu'}\int d\beta \,e^{-W_{\nu'}(\beta)}
	\, \Big[f_{(\nu,\ep)(\nu',+)}^{(0)\eta}(\theta,\beta)f_{\nu',-}^{(0)\eta}(\beta)\n
&&	-f_{(\nu,\ep)(\nu',-)}^{(0)\eta}(\theta,\beta)f_{\nu',+}^{(0)\eta}(\beta)\Big]+ \ldots
\eeqa
and
\beqa\label{LW2}
f^\eta_{(\nu_1,\ep_1)(\nu_2,\ep_2)}(\theta_1, \theta_2)&=&f^{(0)\eta}_{(\nu_1,\ep_1)(\nu_2,\ep_2)}(\theta_1 ,\theta_2)\n
&&+ \sum_\nu\int d\theta\, e^{-W_\nu(\theta)}\Big[f^{(0)\eta}_{ (\nu_1,\ep_1)(\nu,-)}(\theta_1, \theta)f^{(0)\eta}_{ (\nu_2,\ep_2)(\nu,+)}(\theta_2, \theta)\n
&&-f^{(0)\eta}_{ (\nu_1,\ep_1)(\nu,+)}(\theta_1, \theta)f^{(0)\eta}_{ (\nu_2,\ep_2)(\nu,-)}(\theta_2, \theta)\Big]+\ldots.
\eeqa
Then,  the solutions to functional differential equations \eqref{pde1} and \eqref{pde2} can be uniquely fixed order by order, once the zeroth order has been fixed. It can be seen that mixed-state form factors \eqref{mixed-state ff1} and \eqref{mixed-state ff2} in the pure-state limit do reproduce the correct vacuum form factors. So it is sufficient to prove if they are the solution of functional differential equations  (\ref{pde1}) and (\ref{pde2}).

\subsubsection*{\bf{Solution}}

 Consider (\ref{pde1}) first. On the left-hand side, we find 
\beq\label{LHS1}
\frac{\eta\ep f^+_{\nu, \ep}(\theta)}{4\pi i}\frac{1-e^{-2\pi i \eta\nu' \alpha}}{1+e^{-2\pi i \eta\nu' \alpha}e^{-W_{\nu'}(\beta)}}\frac{e^{\frac{\beta-\theta}{2}}}{1+e^{W_{\nu'}(\beta)}}\lt(\frac{\delta_{-\nu,\nu'}}{\cosh(\frac{\theta-\beta}{2})}+\frac{\delta_{\nu,\nu'}}{\sinh(\frac{\theta-\beta}{2})}\rt)
\eeq
and on the right-hand side, using
\beqa
f^{(0)\eta}_{-\nu,-\ep}(\beta)&=&\ep  i e^{(\nu \ep \alpha+1/2)(\beta-\theta)}e^{i\pi\eta\nu \alpha}f^{(0)}_{\nu,\ep}(\beta)\no\,,\\
h^\eta_{\nu',+}(\beta)h^\eta_{\nu',-}(\beta)&=&\frac{1+e^{-W_{\nu'}(\beta)}}{1+e^{-2\pi i \eta\nu' \alpha}e^{-W_{\nu'}(\beta)}}\,,\no
\eeqa
we find
\beq\label{RHS1}
\frac{\eta\ep f^\om_{\nu,\ep}(\theta)}{2\pi}\frac{\sin(\eta\nu'\alpha)e^{-i\pi \eta\nu'\alpha}}{1+e^{-2\pi i \eta\nu' \alpha}e^{-W_{\nu'}(\beta)}}\frac{e^{\frac{\beta-\theta}{2}}}{1+e^{W_{\nu'}(\beta)}}\lt(\frac{\delta_{-\nu,\nu'}}{\cosh(\frac{\theta-\beta}{2})}+\frac{\delta_{\nu,\nu'}}{\sinh(\frac{\theta-\beta}{2})}\rt).
\eeq
 Thanks to the relation
\beq \label{sin}
\sin x=\frac{e^{ix}-e^{-ix}}{2i}\,,
\eeq
\eqref{LHS1} and \eqref{RHS1} are equal.
Then, we consider (\ref{pde2}). On the left-hand side, we find
\beqa\label{LHS2}
&\eta\nu_1\frac{\sin(\pi\alpha)}{8\pi^2}\kappa (1-e^{-2\pi i \eta\nu\alpha})
\lt[\delta_{\ep_1,\ep_2}\delta_{\nu_1,-\nu_2}\lt(\frac{\delta_{\nu_1,\nu}}{\sinh\lt(\frac{\theta_1-\beta}{2}\rt)\cosh\lt(\frac{\theta_2-\beta}{2}\rt)} +\frac{\delta_{-\nu_1,\nu}}{\cosh\lt(\frac{\theta_1-\beta}{2}\rt)\sinh\lt(\frac{\theta_2-\beta}{2}\rt)}\rt)\rt.&\no\\
&\lt.-i\ep_1e^{-i\pi\eta\nu_1\alpha}\delta_{\ep_1,-\ep_2}\delta_{\nu_1,\nu_2}\lt(\frac{\delta_{\nu_1,\nu}}{\sinh\lt(\frac{\theta_1-\beta}{2}\rt)\sinh\lt(\frac{\theta_2-\beta}{2}\rt)} +\frac{\delta_{-\nu_1,\nu}}{\cosh\lt(\frac{\theta_1-\beta}{2}\rt)\cosh\lt(\frac{\theta_2-\beta}{2}\rt)}\rt)\rt]&
\eeqa
where
\beqa
\kappa:=\frac{e^{\ep_1\nu_1(\theta_1-\theta_2)\alpha}}{(1+e^{-2\pi i \eta\nu\alpha}e^{-W_\nu(\beta)})(1+e^{W_\nu(\beta)})}\no
\eeqa
and on the right-hand side, we find
\beqa\label{RHS2}
&\frac{\sin^2(\pi\alpha)}{-4\pi^2}\kappa 
\lt[ie^{-i\pi\eta\nu\alpha}\delta_{\ep_1,\ep_2}\delta_{\nu_1,-\nu_2}\lt(\frac{\delta_{\nu_1,\nu}}{\sinh\lt(\frac{\theta_1-\beta}{2}\rt)\cosh\lt(\frac{\theta_2-\beta}{2}\rt)} -\frac{\delta_{-\nu_1,\nu}}{\cosh\lt(\frac{\theta_1-\beta}{2}\rt)\sinh\lt(\frac{\theta_2-\beta}{2}\rt)}\rt)\rt.&\no\\
&\lt.-\ep_1\delta_{\ep_1,-\ep_2}\delta_{\nu_1,\nu_2}\lt(\frac{\delta_{\nu_1,\nu}e^{-2\pi i\eta\nu_1\alpha}}{\sinh\lt(\frac{\theta_1-\beta}{2}\rt)\sinh\lt(\frac{\theta_2-\beta}{2}\rt)} -\frac{\delta_{-\nu_1,\nu}}{\cosh\lt(\frac{\theta_1-\beta}{2}\rt)\cosh\lt(\frac{\theta_2-\beta}{2}\rt)}\rt)\rt]\,.&
\eeqa
Taking into account the relation (\ref{sin}) again, \eqref{LHS2} and \eqref{RHS2} are equal. It is worth noting that this system of non-linear functional differential equations provides an alternative check of our proposed finite-temperature form factors of $U(1)$ twist fields \eqref{Gibbs1} and \eqref{Gibbs2}.

\subsubsection{General solution as integral-operator kernel}
The non-linear functional differential equations \eqref{pde1} and \eqref{pde2} hold for any local field that can be expressed as normal-ordered exponential of bilinear forms in fermion operators. The exact results \eqref{mixed-state ff1} and \eqref{mixed-state ff2} are just the solution when form factors involved in the differential equations are specialized to those of $U(1)$ twist fields. It is not known that a general solution will also possess the leg-factor structure found. However, following the arguments in \cite{Yixiong}, we can obtain a general solution which is expressed in terms of integral-operator kernels.

 We  consider, without loss of generality,  the primary twist field $\sigma_\alpha^\eta$ \eqref{order field}.
For convenience, we define
\beq
S^\eta:= \sum_{(\nu_1,\ep_1)(\nu_2,\ep_2)}\int d\theta_1d\theta_2 F^\eta_{(\nu_1,\ep_1)(\nu_2,\ep_2)}(\theta_1,\theta_2)D^{\ep_1}_{\nu_1}(\theta_1)D^{\ep_2}_{\nu_2}(\theta_2)
\eeq
and it follows that
\beq
\sigma_\alpha^\eta=\bra \sigma_\alpha \ket \lt(:e^{S^\eta}:\rt)\,.
\eeq
Before we continue, let us introduce a dressing operator $\lU$ in the Liouville space, which is first employed in finite-temperature free Majorana theory \cite{Ben1,Ben2} and then generalized to mixed states \cite{Yixiong}. This operator suggests a particular way of evaluating mixed-state form factors via linear combinations of their pure-state limits \cite{Ben1,Ben2,Yixiong}. In the Dirac theory, we can also define the dressing operator $\lU$:
\beq \label{U}
{\lU} = \exp\lt[\sum_\nu \int d\theta\,\frc{\lZ_{\nu,-}(\theta)\lZ_{\nu,+}(\theta)}{1+e^{W_\nu(\theta)}}\rt].
\eeq
In \cite{Ben1,Ben2,Yixiong}, a normal-ordering operation $\nl\cdot \nl$ in the Liouville space is introduced and this operation brings all Liouville creation operators to the left of Liouville annihilation operators. With this normal-ordering operation, for any usual normal-ordered field $\Or$, we have
\beq \label{U relation}
\Or^\ell |\vac\ket^\rho=\lU \nl \Or^\ell \nl |\vac\ket^\rho\,.
\eeq
See appendix \ref{Proof U} for the proof of relation (\ref{U relation}).

Now, we focus on normalized mixed-state two-particle form factors, for instance, \[{}^{\hspace{14.5mm}\rho}_{(\nu_1,\ep_1)(\nu_2,\ep_2)}\bra\theta_1,\theta_2| \lt(:e^{S^\eta}:\rt)^\ell|\vac\ket^\rho\,.\] Using  (\ref{U relation}), the definition \eqref{liouville mode operators}, the property of Liouville normal-ordering and the fact that $\lU |\vac\ket^\rho=|\vac\ket^\rho$,  we have
\beqa
{}^{\hspace{14.5mm}\rho}_{(\nu_1,\ep_1)(\nu_2,\ep_2)}\bra\theta_1,\theta_2| \lt(:e^{S^\eta}:\rt)^\ell|\vac\ket^\rho&=&{}^{\hspace{14.5mm}\rho}_{(\nu_1,\ep_1)(\nu_2,\ep_2)}\bra\theta_1,\theta_2| \lU\lt(\nl e^{S^{\eta}} \nl\rt)^\ell|\vac\ket^\rho \no\\
&&{}^{\hspace{14.5mm}\rho}_{(\nu_1,\ep_1)(\nu_2,\ep_2)}\bra\theta_1,\theta_2| \lU e^{\t S^\eta} |\vac\ket^\rho \no\\
&&{}^{\hspace{14.5mm}\rho}_{(\nu_1,\ep_1)(\nu_2,\ep_2)}\bra\theta_1,\theta_2|  e^{\lU \t S^\eta \lU^{-1}} |\vac\ket^\rho\no
\eeqa
where 
\beq
\t S^\eta=\sum_{(\nu_1,\ep_1)(\nu_2,\ep_2)}\int d\theta_1 d\theta_2 \frac{F^\eta_{(\nu_1,\ep_1)(\nu_2,\ep_2)}(\theta_1,\theta_2)}{Q^\rho_{\nu_1,-\ep_1}(\theta_2)Q^\rho_{\nu_2,-\ep_2}(\theta_2)}\lZ^\dag_{\nu_1,\ep_1}(\theta_1)\lZ^\dag_{\nu_2,\ep_2}(\theta_2)\,.
\eeq
Thanks to (\ref{UZU^{-1}}), we then have
\beq \label{e^G}
{}^{\hspace{14.5mm}\rho}_{(\nu_1,\ep_1)(\nu_2,\ep_2)}\bra\theta_1,\theta_2| \lt(:e^{S^\eta}:\rt)^\ell|\vac\ket^\rho={}^{\hspace{14.5mm}\rho}_{(\nu_1,\ep_1)(\nu_2,\ep_2)}\bra\theta_1,\theta_2| e^{G^\eta}|\vac\ket^\rho
\eeq
where 
\beqa
G^\eta&=&\sum_{(\nu_1,\ep_1)(\nu_2,\ep_2)}\int d\theta_1 d\theta_2 \,F^\eta_{(\nu_1,\ep_1)(\nu_2,\ep_2)}(\theta_1,\theta_2)\no\\
&&\times \lt( \frac{\lZ^\dag_{\nu_1,\ep_1}(\theta_1)}{Q^\rho_{\nu_1,-\ep_1}(\theta_1)}+\frac{\ep_1 \lZ_{\nu_1,-\ep_1}(\theta_1)}{Q^\rho_{\nu_1,+}(\theta_1)}\rt)
\lt( \frac{\lZ^\dag_{\nu_2,\ep_2}(\theta_2)}{Q^\rho_{\nu_2,-\ep_2}(\theta_2)}+\frac{\ep_2 \lZ_{\nu_2,-\ep_2}(\theta_2)}{Q^\rho_{\nu_2,+}(\theta_2)}\rt)\,.
\eeqa
Now, our problem has been reduced to a question of computing the matrix element of a pure exponential on the right-hand side of  (\ref{e^G}), which can be performed using Bogoliubov transformation, similarly to \cite{Yixiong}.

We construct a basis $b_j,b_j^\dag,c_j,c_j^\dag$ with discrete indices $j=1,2,\ldots,n$, which satisfies canonical anti-commutation relations
\beq
\{b_j,b_k^\dag\}=\{c_j,c_k^\dag\}=\delta_{jk}
\eeq
with other anti-commutators being zero.
We define the vacuum  as $|0\ket$ and the column vector $V$ as \[
V=\lt(b_1,\ldots,c_1;b_1^\dag,\ldots,c_1^\dag\rt)^T ,\quad V^\dag= \lt(b_1^\dag,\ldots,c_1^\dag;b_1,\ldots,c_1\rt) .\]
Concerning the matrix element (\ref{e^G}), we write 
\beq \label{trace}
\bra0|VV^\dag e^{V^\dag J V}|0\ket=\lim_{\beta\to \infty}\Tr(e^{-\beta N}VV^\dag e^{V^\dag J V})\,,\quad N=\frac 1 2 V^\dag \sigma_z V
\eeq
where $\sigma_z$ is pauli matrix, $J$ is a general 2-block by 2-block matrix and $\beta$ is a real parameter. 
When evaluating the trace in (\ref{trace}), we can move $V$ along one cycle by using cyclic property of the trace and commutation relations
\beq
[N,V]=-\sigma_zV,\quad [V^\dag J V,V]=MV,\quad M=\sigma_x J^T \sigma_x-J,\quad \{V,V^\dag\}=I
\eeq
where $\sigma_x$ is pauli matrix and $I$ represents the identity matrix. Then we get the relation
\beq
\lt(1+e^\beta \sigma_z e^M\rt)\Tr\lt(e^{-\beta N}VV^\dag e^{V^\dag J V}\rt)=e^\beta \sigma_z e^M
\Tr\lt(e^{-\beta N}e^{V^\dag J V}\rt)\,.
\eeq
Take the limit $\beta \to \infty$ and consider only the divergent terms proportional to $e^\beta$. We have
\beq
\left(\begin{array}{cc} (e^M)_{11} &  (e^M)_{12}\\ 0 &  0\end{array}\right)\bra 0|\left(\begin{array}{cc} (VV^\dag)_{11} &  (VV^\dag)_{12}\\ (VV^\dag)_{21} &  (VV^\dag)_{22}\end{array}\right)e^{V^\dagger JV}|0\ket=\left(\begin{array}{cc} (e^M)_{11} &  (e^M)_{12}\\ 0 &  0\end{array}\right)\bra 0|e^{V^\dagger JV}|0\ket\no\,.
\eeq
Using rules of matrix product and $\bra 0| (VV^\dag)_{22}e^{V^\dag J V}|0\ket=0$, we find
\beq
\frac{\bra 0| (VV^\dag)_{12}e^{V^\dag J V}|0\ket}{\bra 0|e^{V^\dagger JV}|0\ket}=\lt( (e^M)_{11}\rt)^{-1}(e^M)_{12}
\eeq
which we are interested in for later computation.

Now, we can apply these techniques to our  two-particle form factor problem. It is worth noting that these techniques are only valid for Fock space based on canonical anti-commutation algebra. So we have to define new Liouville mode operators
\beq
\lb^\dag_{\nu,\ep}(\theta)=\frac{\sqrt {Q^\rho_{\nu,+}(\theta)}}{Q^\rho_{\nu,-\ep}(\theta)}\lZ^\dag_{\nu\ep}(\theta)\,\,,\quad 
\lb_{\nu,\ep}(\theta)=\frac {1}{\sqrt{Q^\rho_{\nu,+}(\theta)}}\lZ_{\nu,\ep}(\theta)
\eeq
so that they satisfy anti-commutation relation
\beqa
\{\lb_{\nu,\ep}(\theta),\lb^\dag_{\nu',\ep'}(\theta')\}&=&\delta_{\nu,\nu'}\delta_{\ep,\ep'}\delta(\theta-\theta')\,,\no\\
\{\lb_{\nu,\ep}(\theta),\lb_{\nu',\ep'}(\theta')\}&=&\{\lb^\dag_{\nu,\ep}(\theta),\lb^\dag_{\nu',\ep'}(\theta')\}=0\no\,.
\eeqa
We then see
\beqa
G^\eta&=&\sum_{(\nu_1,\ep_1)(\nu_2,\ep_2)}\int \frac{d\theta_1 d\theta_2}{\sqrt{Q^\rho_{\nu_1,+}(\theta_1)Q^\rho_{\nu_2,+}(\theta_2)}} \,F^\eta_{(\nu_1,\ep_1)(\nu_2,\ep_2)}(\theta_1,\theta_2)\no\\
&&\times \lt( \lb^\dag_{\nu_1,\ep_1}(\theta_1)+\ep_1\lb_{\nu_1,-\ep_1}(\theta_1)\rt)
\lt( \lb^\dag_{\nu_2,\ep_2}(\theta_2)+\ep_2\lb_{\nu_2,-\ep_2}(\theta_2)\rt)\,.
\eeqa
Identifying $G^\eta=V^\dag J V$, and considering the index $(\nu,\ep)$, we can write down matrix $M$ in the 8 by 8 form 

\beq
\renewcommand{\arraystretch}{1.5}
	M = \mato{cccccccc} 
     x_+ & 0 & 0 & -y & 0 & -x_+ & -y & 0  \\
     0 & x_+^{t} & h & 0 & x_+^t & 0 & 0 & -h  \\
     0 & y^t & x_- & 0 & y^t & 0 & 0 & -x_b  \\
     -h^t & 0 & 0 & x_-^t & 0 & h^t & x_-^t & 0  \\
     0 & -x_+^t & -h & 0 & -x_+^t & 0 & 0 & h  \\
     x_+ & 0 & 0 & -y & 0 & -x_+ & y & 0  \\
     h^t & 0 & 0 & -x_-^t & 0 & -h^t & -x_-^t & 0  \\
     0 & y^t & x_- & 0 & y^t & 0 & 0 & -x_-  \matf
\eeq
where the integral operators $x_+$, $x_-$, $y$ and $z$ have kernels
\beqa
	x_+(\theta_1,\theta_2)
	&=& \frc{2F^\eta_{(+,+)(+,-)}(\theta_1,\theta_2)
	}{\sqrt{Q^\rho_{+,+}(\theta_1)Q^\rho_{+,+}(\theta_2)}} \n
     	x_-(\theta_1,\theta_2)
	&=& \frc{2F^\eta_{(-,+)(-,-)}(\theta_1,\theta_2)
	}{\sqrt{Q^\rho_{-,+}(\theta_1)Q^\rho_{-,+}(\theta_2)}} \n
	y(\theta_1,\theta_2) &=&
	\frc{2F^\eta_{(+,+)(-,+)}(\theta_1,\theta_2)
	}{\sqrt{Q^\rho_{+,+}(\theta_1)Q^\rho_{-,+}(\theta_2)}} \n
	h(\theta_1,\theta_2) &=& \frc{2F^\eta_{(+,-)(-,-)}        (\theta_1,\theta_2)
	}{\sqrt{Q^\rho_{+,+}(\theta_1)Q^\rho_{-,+}(\theta_2)}}\,.
	 \no
\eeqa
and $t$ represents matrix transpose. Thanks to the notion that $M$ is nilpotent, namely $M^2=0$, we can express its exponential simply as \[
e^M=1+M \,. \] 
Direct calculations show that
\beq
	R:=\lt( (e^M)_{11}\rt)^{-1}(e^M)_{12} = \mato{cccc}
	0 & g & k & 0 \\ -g^t & 0 & 0 & l \\
	-k^t & 0 & 0 & m \\ 0 & -l^t & -m^t & 0
       \matf
\eeq
where
\beqa
g&=&\lt[ x_++ I-y\lt(x_-^t+I\rt)^{-1}h^t\rt]^{-1}-I\,,\quad k=\lt[h^t-\lt(x_-^t+I\rt)y^{-1}(x_++I) \rt]^{-1}\\
m&=&\lt[ x_-+I-y^t\lt(x_+^t+I\rt)^{-1}h\rt]^{-1}-I  \,,\quad\,
l=\lt[y^t-\lt(x_-+I\rt)h^{-1}(x_+^t+I) \rt]^{-1}\,.
\eeqa
We are thus led to conclude that two-particle form factors can be given via the kernel of integral-operator $R$,
\beq
   {}^{\hspace{14.5mm}\rho}_{(\nu_1,\ep_1)                                                                                                            
   (\nu_2,\ep_2)}\bra\theta_1,\theta_2| \lt(:e^{S^\eta}:\rt)^\ell|  
    \vac\ket^\rho=\sqrt{Q^\rho_{\nu_1,+}(\theta_1)Q^\rho_{\nu_2,+} 
    (\theta_2)}\,R_{(\nu_1,\ep_1)(\nu_2,\ep_2)}(\theta_1,\theta_2)
\eeq
where $(+,+)$, $(+,-)$ are on the first two rows/columns and $(-,+)$, $(-,-)$ are on the second.

\subsection{Mixed-state two-point correlation functions of twist fields}

In the Ising model, a correlation function of twist fields  comes from the scaling limit of the correlation function of corresponding spin operators in the quantum Ising chain (see for instance \cite{Drouffe11111}). The choice of a direction for the branch cut must be kept the same for each twist fields inside the correlation function, due to the Jordan-Wigner transformation \cite{Lieb61,Katsura62,JordanWigner} in which the Pauli spin matrices are written as infinite products of fermion operators starting at the matrix's site and going in a fixed direction. Since the Dirac theory can be seen as a doubled Ising model, we consider again mixed-state two-point correlation functions of $U(1)$ twist fields with branch cuts going towards the same direction (towards the right):
\[
\bra\sigma^+_\alpha(x,t)\sigma^+_{\alpha'}(0,0)\ket_\rho\quad \text{and}\quad \bra\sigma^+_{\alpha\pm1,\alpha}(x,t)\sigma^+_{\alpha'\mp1,\alpha'}(0,0)\ket_\rho\,.
\]

With mixed-state form factors at hand, we can obtain a series of expression for mixed-state two-point correlation functions of local fields $\Or_1$ and $\Or_2$, using the resolution of the identity (\ref{resolution})
\beqa\label{2points for NIF}
 &&\lefteqn{\bra \Or_1(x,t) \Or_2(0,0)\ket_\rho}\n
&=&\sum_{N=0}^{\infty}\sum_{\nu_1,\ldots,\nu_N} \sum_{\ep_1,\ldots,\ep_N}\int \frac{d\theta_1 \cdots d\theta_N}{N!}\lt[\frac{e^{\sum_{j=1}^N \lt(i\ep_jp_{\theta_j} x - i\ep_jE_{\theta_j} t\rt)}}{\prod_{j=1}^{N}\lt(1+e^{-\ep_j W_{\nu_j}(\theta_j)}\rt)}\rt.\n
&&\lt.f^{\rho;\Or_1}_{(\nu_1,\ep_1)\ldots(\nu_N,\ep_N)}(\theta_1,\ldots,\theta_N) f^{\rho; \Or_2}_{(\nu_N,-\ep_N)\ldots(\nu_1,-\ep_1)}(\theta_N,\ldots,\theta_1) \rt] \,.
\eeqa
This expression is expected to hold for any non-interacting field $\Or$ whose form factors are zero  for large enough numbers of particles so that the above series truncates. 

However, some analysis of the integrals in \eqref{2points for NIF} needs to be performed. If the value of $W_{\nu_j}(\theta_j)$ increases as $|\theta_j|\to\infty$, then the integral over $\theta_j$ is convergent for $\ep_j=-$. However, the integral over $\theta_j$ is in general not convergent for $\ep_j=+$. In analogy with the standard $i0^+$ prescription for  correlation functions in the context of QFT, assuming $W_{\nu_j}(\theta_j)$ grows like, or faster than $e^{\alpha\cosh\theta_j}$ for some $\alpha>0$ as $|\theta_j|\to\infty$, we can make both cases $\ep_j=\pm$ convergent  by replacing $t$ with $t-i0^+$. With this prescription, the correlation function is seen as the boundary value, at $t\in\R$, of a function of $t$ analytic on some neighborhood of $\R$ in the region ${\rm Im}(t)<0$. 

In fact, we can make this boundary value  finite at space-like distances ($x^2>t^2$) for any $W_{\nu_j}(\theta_j)$ as long as  $W_{\nu_j}(\theta_j)$ is analytic on neighborhoods of $(K,\infty)$ and $(-\infty,-K)$, for $K>0$ large enough.  To see this, assuming without loss of generality that $x>0$, we shift the contours as $\theta_j \mapsto \theta_j+\ep_j i0^+$ in the region $\lt|{\rm Re}(\theta_j)\rt|>K$, so that $\theta_j$ remains in the analyticity region of $W_{\nu_j}(\theta_j)$.  It turns out that this boundary value with integrals on the shifted contours is indeed finite at space-like distances ($x^2>t^2$).

For twist fields, the form factor expansion is infinite, as these fields have non-zero form factors for arbitrary large numbers of particles. However, the resulting infinite series (\ref{2points for NIF}) is not the correct representation for mixed-state  two-point functions of twist fields. The form factor expansion needs to be modified in various ways, because of the branch cuts emanating from twist fields as expressed in the twist condition. We present the arguments for these modifications and then propose the form factor expansions for two-point functions of $U(1)$ twist fields in mixed states. From now on, we take without loss of generality $x>0$ and only consider the space-like region $x^2>t^2$.

\subsubsection{Three modifications}

First, if  the general expansion $\sum_s {}^\rho \bra \vac| \Or_1(x,t)^\ell |s\ket^\rho\,{}^\rho\bra s | \Or_2(0,0)^\ell |\vac\ket^\rho$ is a large distance expansion, then $|s\ket^\rho\,{}^\rho\bra s | $ should be interpreted as intermediate states over the region between the fields $\Or_1(x,t)^\ell$ and $ \Or_2(0,0)^\ell$, and the vacuum states $^\rho \bra \vac|$ and $|\vac\ket^\rho$ should represent what is happening on the far right and left respectively. It has been argued in \cite{Ben1,Ben2}, via the comparison between finite-temperature form factors and form factors on the circle, that  the intermediate states must lie in a region which is not affected by the branch cuts of twist fields. This means that we have to obtain form factor expansions where no cut is present in the region between $0$ and $x$. Fortunately, the unitary operator $Z$ can help us achieve this:
\beqa
\bra\sigma^+_\alpha(x,t)\sigma^+_{\alpha'}(0,0)\ket_\rho&=&\bra\sigma^+_\alpha(x,t)\sigma^-_{\alpha'}(0,0)\ket_{\rho^\sharp}\no\\
 \bra\sigma^+_{\alpha\pm1,\alpha}(x,t)\sigma^+_{\alpha'\mp1,\alpha'}(0,0)\ket_\rho&=&\bra\sigma^+_{\alpha\pm1,\alpha}(x,t)\sigma^-_{\alpha'\mp1,\alpha'}(0,0)\ket_{\rho^\sharp}
\eeqa
where we relate $\sigma_{\alpha'}^+$ and $\sigma^+_{\alpha'\mp1,\alpha'}$ to $\sigma_{\alpha'}^-$ and $\sigma^-_{\alpha'\mp1,\alpha'}$ via operator $Z$ respectively, and we get the twisted density matrix 
\vspace{0.1cm}
$\rho^\sharp:=Z^{-1}\rho$ with $W^\sharp_\pm(\theta)=W_\pm(\theta)\pm2\pi i \alpha'$.
Here, two-point functions \[\bra\sigma^+_{\alpha\pm1,\alpha}(x,t)\sigma^+_{\alpha'\pm1,\alpha'}(0,0)\ket_\rho\] are 
not considered, as they are all zero, due to the fact that twist fields $\sigma^\eta_{\alpha+1,\alpha}$ have non-zero one-particle form factors $f^{\rho;\sigma^\eta_{\alpha+1,\alpha}}_{\nu,\ep}(\theta)$ only for $\nu=\ep$ while twist fields $\sigma^\eta_{\alpha-1,\alpha}$ have non-zero one-particle form factors $f^{\rho;\sigma^\eta_{\alpha-1,\alpha}}_{\nu,\ep}(\theta)$ only for $\nu=-\ep$.

Second, the insertion of a twist field inside mixed-state correlation functions or traces will affect one of the vacuum sectors in the correspondence to vacuum expectation values in the quantization on the circle and hence gives rise to the free energy difference between the sector where a cut lies and the one where no cut lies. This property can be expressed via
\beq\label{transcov}
	{}^\rho\bra \vac|\omega^\eta(x,t)^\ell|\theta_1,\ldots, \theta_N\ket^\rho_{(\nu_1,\ep_1)\ldots(\nu_N,\ep_N)}
	=
	e^{\eta x{\cal E}}
	e^{\sum_{j=1}^N \lt(i\ep_jp_{\theta_j} x - i\ep_jE_{\theta_j} t\rt)}
	f^{\rho;\omega^\eta}_{(\nu_1,\ep_1)\ldots(\nu_N,\ep_N)}(\theta_1,\ldots,\theta_N)
\eeq
for both twist fields $\om=\sigma_\alpha$ and $\om=\sigma_{\alpha\pm1,\alpha}$, with $\cal E$ the free energy deficit:
\beq
{\cal E}=\sum_{\nu=\pm}\int\frac{d\theta}{2\pi}m\cosh\theta \log\lt( \frac{1+e^{-W_{\nu}(\theta)}}{1+e^{-2\pi i \nu \alpha}e^{-W_{\nu}(\theta)}}\rt)\,.
\eeq
 We denote by ${\cal E}^\sharp$ the free energy deficit associated to $W_\nu^\sharp$ and we have
\beq
{\cal E}^\sharp=-{\cal E}\,.
\eeq

Finally, mixed-state form factors of twist fields are not entire functions of the rapidities but distributions defined as boundary values of analytic functions with no colliding rapidities. We have to shift the contour towards the analytic region for the purpose of obtaining a well-defined form factor expansion. Therefore, we need to further require that $W_\nu(\theta)$ be analytic on a neighborhood of $\R$.

\subsubsection{Two-point correlation functions}
According to the modifications stipulated above, we propose the followings:
\begin{propo} \label{conjexp}
With $W_\nu(\theta)$ analytic on a neighborhood of $\theta\in\R$, we have
\begin{eqnarray} \label{correlation1}
&&\lefteqn{\bra\sigma^{+}_\alpha(x,t) \sigma^{+}_{\alpha'}(0,0)\ket_\rho}\\
&=&
e^{-x{\cal E}}\sum_{N=0}^{\infty}\sum_{\nu_1,\ldots,\nu_N} \sum_{\ep_1,\ldots,\ep_N}\int \frac{d\theta_1 \cdots d\theta_N}{N!}\lt[\frac{e^{\sum_{j=1}^N \lt(i\ep_jp_{\theta_j} x - i\ep_jE_{\theta_j} t\rt)}}{\prod_{j=1}^{N}\lt(1+e^{-\ep_j\nu_j 2 \pi i\alpha}e^{-\ep_j W_{\nu_j}(\theta_j)}\rt)}\rt.\n
&&\lt.f^{\rho^\sharp;\sigma_\alpha^{+}}_{(\nu_1,\ep_1)\ldots(\nu_N,\ep_N)}(\theta_1,\ldots,\theta_N) f^{\rho^\sharp; \sigma^{-}_{\alpha'}}_{(\nu_N,-\ep_N)\ldots(\nu_1,-\ep_1)}(\theta_N,\ldots,\theta_1) \rt]
\no
\end{eqnarray}
and
\begin{eqnarray} \label{correlation2}
&&\lefteqn{\bra\sigma^+_{\alpha\pm1,\alpha}(x,t)\sigma^+_{\alpha'\mp1,\alpha'}(0,0)\ket_\rho}\\
&=&
e^{-x{\cal E}}\sum_{N=0}^{\infty}\sum_{\nu_1,\ldots,\nu_N} \sum_{\ep_1,\ldots,\ep_N}\int \frac{d\theta_1 \cdots d\theta_N}{N!}\lt[\frac{e^{\sum_{j=1}^N \lt(i\ep_jp_{\theta_j} x - i\ep_jE_{\theta_j} t\rt)}}{\prod_{j=1}^{N}\lt(1+e^{-\ep_j\nu_j 2 \pi i\alpha}e^{-\ep_j W_{\nu_j}(\theta_j)}\rt)}\rt.\n
&&\lt.f^{\rho^\sharp;\sigma^+_{\alpha\pm1,\alpha}}_{(\nu_1,\ep_1)\ldots(\nu_N,\ep_N)}(\theta_1,\ldots,\theta_N) f^{\rho^\sharp;\sigma^-_{\alpha'\mp1,\alpha'}}_{(\nu_N,-\ep_N)\ldots(\nu_1,-\ep_1)}(\theta_N,\ldots,\theta_1) \rt]\,.
\no
\end{eqnarray}

\end{propo}
Form factor expansions (\ref{correlation1}) and (\ref{correlation2}) are infinite since twist fields have non-zero mixed-state form factors for arbitrarily large number of particles. In order to have convergent integrals in our expansions, we shift the $\theta_j$ contours in (\ref{correlation1}) and (\ref{correlation2}) by $i\ep_j \zeta$ for $\zeta>0$ small enough in such a way that $\theta_j$ remains in the analyticity region of $W_\nu(\theta_j)$ and of  the form factors involved. 

The large distance leading behaviour of the two-point functions (\ref{correlation1}) and (\ref{correlation2}) are determined by singularities of the function $\lt(1+e^{-\ep\nu 2 \pi i\alpha}e^{-\ep W_{\nu}(\theta)}\rt)^{-1}$. Among these singularities, we denote by $\theta^\star$ the one with the smallest value of $|{\rm Im}(\sinh\theta)|$. we have
\beqa \label{xiangbuchulai}
\bra\sigma^{+}_\alpha(x,t) \sigma^{+}_{\alpha'}(0,0)\ket_\rho&=&\bra \sigma^+_\alpha\ket_{\rho^\sharp} \,  \bra \sigma^-_{\alpha'}\ket_{\rho^\sharp}  \,e^{-x{\cal E}}\lt(1 +
	O\lt(e^{- mx |{\rm Im}(\sinh \theta_1^\star)|- mx |{\rm Im}(\sinh \theta_2^\star)|}\rt) \rt)\n
\bra\sigma^+_{\alpha\pm1,\alpha}(x,t)\sigma^+_{\alpha'\mp1,\alpha'}(0,0)\ket_\rho&=&O\lt(e^{-x{\cal E}-mx|{\rm Im}(\sinh \theta^\star)|}\rt)
\eeqa
where exponential decays include possible algebraic or other non-exponential factors in $mx$ determined by the type of singularities.

\section{Application: R$\acute{\text{e}}$nyi entropy for integer $n$ in the Ising model}

The R$\acute{\text{e}}$nyi entropy can be used to evaluate the bipartite entanglement entropy which is a measure of quantum entanglement \cite{Bennett}. For the definition of the entanglement entropy, consider a composite quantum system with Hilbert space $\mH=\mH_A \otimes \mH_B$ in a ground state $|\text{gs}\ket$. The entanglement entropy $S_A$ is the von Neumann entropy associated to the reduced density matrix $\rho_A$ of the subsystem $A$:
\beq
S_A=-\Tr_{\mH_A}\lt(\rho_A\log(\rho_A)\rt),\quad \rho_A=\Tr_{\mH_B}\lt(|\text{gs}\ket \bra \text{gs}|\rt)\,.
\eeq
Using the ``replica trick'' \cite{Holzhey,Calabrese}, the bipartite entanglement entropy can be obtained as the limit $n\rightarrow 1$ of the R$\acute{\text{e}}$nyi entropy for positive real $n$:
\beq
S_A=\lim_{n\rightarrow1} S_A^{(n)},\quad  S_A^{(n)}=\frac{1}{1-n}\log \Tr_{\mH_A}\rho_A^n\,.
\eeq
 It has been demonstrated in \cite{Holzhey} that the R$\acute{\text{e}}$nyi entropy for integer $n$ is related to the partition function on a multi-sheeted Riemann surface with branch points. The authors in \cite{Knizhnik,Cardy} introduced the so-called branch-point twist fields which correspond to branch points so that their correlation functions are the partition functions on multi-sheeted Riemann surfaces. Therefore, the problem of calculating the biparticle entanglement entropy is reduced to the evaluation of the two-point correlation function of the branch-point twist fields. In \cite{Cardy}, the two-point correlation function of the branch-point twist fields were computed by exploiting the factorized scattering method for integrable models of QFT.

In this section, we consider the composite system mentioned above in mixed states in the Ising model. In this case, the R$\acute{\text{e}}$nyi entropy for integer $n$ is related with the mixed-state two-point correlation function of the branch-point twist fields in the $n$-copy Ising model. Thanks to the fact that the branch-point twist fields in the $n$-copy Ising model, for even values of $n$, have explicit representations in terms of the $n$ independent $U(1)$ twist fields in the $n$-copy free massive Dirac theory (see Appendix B of \cite{Cardy}), our result \eqref{correlation1} can be applied directly to the evaluation of the mixed-state two-point function of the branch-point twist fields in the $n$-copy Ising model.  Taking analytic continuation in $n$ of the R$\acute{\text{e}}$nyi entropy and computing the result at $n=1$ could in principle give the mixed-state bipartite entanglement entropy. However, this is beyond the scope of this paper.

\subsection{Branch-point twist fields in the $n$-copy Ising model}

The branch-point twist fields originally arose in the computation of the leading large-distance correction to the bipartite ground-state entanglement entropy in the context of massive integrable quantum field theories with diagonal scattering matrices \cite{Cardy}.
Let us commence with a brief review of the branch point twist fields in the Ising model. Consider a model which is formed by $n$ independent copies of the Ising model. Particles on different copies do not interact. The lagrangian density of this $n$-copy Ising model is the sum of the langrangian density of every copy and it can be written as
\beq
\liou^{(n)}[\psi_1, \dots, \psi_n](x)=\liou[\psi_1](x)+\dots+\liou[\psi_n](x)
\eeq
where $\psi_i$ is the free Majorana field on the $i^{\text{th}}$ copy of the model. It is obvious that this model possesses a $\mathbb{Z}_n$ symmetry under cyclic exchange of the copies:
\beq
\liou^{(n)}[g\psi_1, \dots, g\psi_n](x)=\liou^{(n)}[\psi_1, \dots, \psi_n](x)
\eeq
where $g$ is the transformation that permutes the copy numbers cyclically:
\beq
g\psi_i=\psi_{i+1}
\eeq
with $i=1,\cdots,n,n+1\equiv1$. The branch-point twist field $\mathrm{\mathcal{T}}$ is the twist field  associated to the symmetry $g$. Denoting by $\T^+$ and $\T^-$ the twist fields with branch cuts on the right and on the left, respectively, their equal-time exchange relations with respect to the free Majorana fields are:
\beq\label{branch-point exchange1}
      \psi_i(x) \T^{\mbox{\small$\eta$}}(0)= \lt\{
\begin{array}{ll} \T^{\mbox{\small$\eta$}}(0) \lt(\delta_{\eta+} \psi_{i}(x)+\delta_{\eta-}\psi_{i+1}(x) \rt) & \quad(x<0) \\
\T^{\mbox{\small$\eta$}}(0)\lt(\delta_{\eta-} \psi_{i}(x)+\delta_{\eta+}\psi_{i+1}(x) \rt) &  \quad(x>0)\ea \rt.
\eeq
with $\eta=\pm$.
In addition, we can define another branch-point twist field $\tilde\T$ which is associated to the symmetry $g^{-1}$ under the opposite cyclic exchange $g^{-1}\psi_{i}=\psi_{i-1}$ with $i-n=i$. Similarly, we denote by $\tilde\T^+$ and $\tilde\T^-$ the twist fields with branch cuts on the right and on the left, respectively. They also have the equal-time exchange relations with respect to the free Majorana fields:
\beq\label{branch-point exchange2}
      \psi_i(x) \tilde\T^{\mbox{\small$\eta$}}(0)= \lt\{
\begin{array}{ll} \tilde\T^{\mbox{\small$\eta$}}(0) \lt(\delta_{\eta+}\psi_{i}(x)+\delta_{\eta-}\psi_{i-1}(x)\rt) & \quad(x<0) \\
 \tilde\T^{\mbox{\small$\eta$}}(0) \lt(\delta_{\eta-}\psi_{i}(x)+\delta_{\eta+}\psi_{i-1}(x)\rt)  &  \quad(x>0)\ea \rt.
\eeq
with the identification $i-n=i$. It is implied from \eqref{branch-point exchange1} and \eqref{branch-point exchange2} that the branch-point twist field $\tilde\T^{\mbox{\small$\eta$}}$ is the Hermitian conjugate of the branch-point twist field $\T^{\mbox{\small$\eta$}}$:
\beq \label{conjugate relation}
\tilde\T^{\mbox{\small$\eta$}}=(\T^{\mbox{\small$\eta$}})^\dag\,.
\eeq
These twist fields are spinless and they are primary fields with the lowest possible scaling dimension (they have the same scaling dimension) \cite{Knizhnik}:
\beq \label{dimension}
d_n=\frac{1}{24}(n-\frac{1}{n})\,.
\eeq
The branch-point twist fields $\T^{\mbox{\small$\eta$}}$ and $\tilde\T^{\mbox{\small$\eta$}}$, defined by \eqref{branch-point exchange1} and \eqref{branch-point exchange2} respectively, can be uniquely fixed by the requirement that  they have the lowest scaling dimension given by \eqref{dimension} and they are invariant under all symmetries of the $n$-copy Ising model which commute with $g$. 

\subsection{Explicit representation of the branch-point twist fields in the $n$-copy Ising model in terms of the $U(1)$ twist fields in the $n$-copy Dirac theory}

As we mentioned above, there exists a relation between the branch-point twist fields in the $n$-copy Ising model and the $U(1)$ twist fields in the $n$-copy Dirac theory. 
To see this relation, we construct an $n$-copy free Dirac fermion model by doubling the  $n$-copy Ising model. We denote the fundamental real Majorana fermion fields for each $n$-copy Ising model by $\psi_{a,j},\bar\psi_{a,j}$ and  $\psi_{b,j},\bar\psi_{b,j}$ for $j=1,\cdots,n$, respectively, and the fundamental Dirac spinor fermion field by \[\Psi_j=\lt(\begin{array}{c} \Psi_{R,j}
 \\ \Psi_{L,j} \end{array}\rt)\,.\] Then we have the identification:
\beq
\Psi_{R,j}=\frac{1}{\sqrt{2}}(\psi_{a,j}+i\psi_{b,j}),\quad\quad \Psi_{L,j}=\frac{1}{\sqrt{2}}(\bar\psi_{a,j}-i\bar\psi_{b,j})\,.
\eeq
In order for different copies of the Dirac fermions to anti-commute with each other, we define a new basis 
\beq \label{ac basis}
\lt(\ba{c}\Psi^{\text{ac}}_1 \\
\vdots\\
\Psi^{\text{ac}}_n\ea\rt)
\eeq with scattering matrix $-1$ among different copies. Accordingly, the branch-point twist field in the $n$-copy Dirac theory, which we denote by $\TD^\eta$, has  modified exchange relations with respect to the Dirac fermions:
\beq
      \Psi^{\text{ac}}_j(x) \TD^+(0)= \lt\{
\begin{array}{lll} \TD^+(0)\Psi^{\text{ac}}_j(x) & \quad\mbox{for $j=1,\cdots,n$}  & \quad(x<0) \\
\TD^+(0)\Psi^{\text{ac}}_{j+1}(x) & \quad\mbox{for $j=1,\cdots,n-1$} &  \quad(x>0) \\
-\TD^+(0)\Psi^{\text{ac}}_{1}(x) & \quad\mbox{for $j=n$} &  \quad(x>0)
\ea\rt.
\eeq
and 
\beq
      \Psi^{\text{ac}}_j(x) \TD^-(0)= \lt\{
\begin{array}{lll} \TD^-(0)\Psi^{\text{ac}}_j(x) & \quad\mbox{for $j=1,\cdots,n$}  & \quad(x>0) \\
\TD^-(0)\Psi^{\text{ac}}_{j+1}(x) & \quad\mbox{for $j=1,\cdots,n-1$} &  \quad(x<0) \\
-\TD^-(0)\Psi^{\text{ac}}_{1}(x) & \quad\mbox{for $j=n$} &  \quad(x<0)\,.
\ea\rt.
\eeq
Then, we diagonalise the branch-point twist fields in the $n$-copy Dirac theory by  performing a $SU(n)$ transformation of the basis \eqref{ac basis} and the new basis after this transformation can be considered as $n$ independent Dirac fermions. In this new basis, the branch-point twist fields can be written as a product of $U(1)$ twist fields acting on these independent Dirac fermions from different copies:
\beq \label{relation B and U}
\T^\eta_{\text{Dirac}}=\prod_{k} \sigma^\eta_{(k,\alpha_k)}
\eeq 
where $k=1,\ldots,n$ represents the copy number and $\alpha_k=\frac{2k-n-1}{2n}$ is associated with the $U(1)$ element $e^{2\pi i\alpha_k}$, for $\alpha_k \in [-1,1]$. It is worth mentioning that the relation \eqref{relation B and U} is only valid in the case of even $n$. For $n$ odd, the dimension of the branch-point twist field constructed from this factorisation relation does not agree with \eqref{dimension} which is predicted by the conformal field theory. More details of the derivation of \eqref{relation B and U} can be found in Appendix B of \cite{Cardy}. On the other hand, from the point of view that the $n$-copy Dirac theory is a doubled $n$-copy Ising model, the branch-point twist field $\T_{\text{Dirac}}^\eta$ has the relation 
\beq \label{Tab}
\T_{\text{Dirac}}^\eta=\T_a^\eta \otimes \T_b^\eta
\eeq
where $
\T_a^\eta$ and $\T_b^\eta$ are the branch-point twist fields in the copies $a$ and $b$ of the $n$-copy Ising model, respectively.

\subsection{R$\acute{\text{e}}$nyi entropy for integer $n$}

According to the arguments in \cite{Cardy}, the R$\acute{\text{e}}$nyi entropy for integer $n$ in the Ising model can be written in terms of the mixed-state two-point correlation function of the branch-point twist fields ( without loss of generality we consider only the branch-point twist fields with cuts going towards the right):
\beq\label{renyi n}
S_A^{(n)}=\frac{1}{1-n}\log\lt[\varep^{2d_n} \tilde Z_n \big\bra \T(x,0)\tilde\T(0,0)\big\ket_{\rhoni}\rt]
\eeq
where $\tilde Z_n$ is an $n$-dependent non-universal normalisation constant with $\tilde Z_1=1$, $\varep$ is a short-distance cutoff which is chosen in such a way that $d\tilde Z_n/dn=1$, $d_n$ is the scaling dimension \eqref{dimension}, and $\rhoni$ represents the density matrix of the $n$-copy Ising model. 

Let us now evaluate the two-point function $\big\bra \T(x,0)\tilde\T(0,0)\big\ket_{\rhoni}$.
 We first define the density matrix $\rhond$ for the $n$-copy Dirac theory:
\beq
\rhond=\rho^1_{\text{D}}\otimes\cdots\otimes\rho_{\text{D}}^n
\eeq
where $\rho^i_{\text{D}}$ is the density matrix in the $i^{\text{th}}$-copy of the model. Then, in light of \eqref{relation B and U}, the two-point correlation function of branch-point twist fields $\TD(x,0)$ and $\tTD(0,0)$ in the $n$-copy Dirac theory in mixed states can be written as
\beqa \label{mixed correlation 2T 1}
&&\big\bra \TD^+(x,0) \tTD^+(0,0)\big\ket_{\rhond}\n
&=&\frac{{\Large\text{\Tr}}_{\mH^{(n)}_\td}\lt(\rhond\prod_{m=1}^n \sigma_{(m,\alpha_m)}^+(x,0)\prod_{p=1}^n \sigma_{(p,-\alpha_p)}^+(0,0)\rt)}{{\Large\text{\Tr}}_{\mH^{(n)}_\td}\lt(\rhond\rt)}\n
&=&\frac{\ltr_{\mH^1_\td}\lt(\rho^1_{\text{D}}\,\sigma_{(1,\alpha_1)}^+(x,0)\sigma_{(1,-\alpha_1)}^+(0,0)\rt)}{\ltr_{\mH^1_\td}\lt(\rho^1_{\text{D}}\rt)}\cdots \frac{\ltr_{\mH^n_\td}\lt(\rho^n_{\text{D}}\,\sigma_{(n,\alpha_n)}^+(x,0)\sigma_{(n,-\alpha_n)}^+(0,0)\rt)}{\ltr_{\mH^n_\td}\lt(\rho^n_{\text{D}}\rt)}\n
&=&\big\bra \sigma_{(1,\alpha_1)}^+(x,0)\sigma_{(1,-\alpha_1)}^+(0,0)\ket_{\rho^1_{\text{D}}}\cdots \bra \sigma_{(n,\alpha_n)}^+(x,0)\sigma_{(n,-\alpha_n)}^+(0,0)\big\ket_{\rho^n_{\text{D}}}
\eeqa
where $\mH^{(n)}_\td$ is the Hilbert space of the $n$-copy Dirac theory and $\mH^i_\td$ is the Hilbert space of the $i^{\text{th}}$-copy of it, and where in the first step we used relations \eqref{conjugate relation} and \eqref{sigma HC}. On the other hand, we can define the density matrix $\rhond$ in another way:
\beq
\rhond=\rho^{(n)}_{\text{I}_a} \otimes \rho^{(n)}_{\text{I}_b}
\eeq
where $\rhonia$ and $\rhonib$ are the density matrices in the copies $a$ and $b$ of the $n$-copy Ising model, respectively. Using the relation \eqref{Tab}, we have
\beqa \label{mixed correlation 2T 2}
&&\big\bra \TD^+(x,0) \tTD^+(0,0)\big\ket_{\rhond}\n
&=& \frac{\ltr_{\mH^{(n)}_{\td}}\lt(\rho^{(n)}_\td\T_a^+(x,0)\T_b^+(x,0) \tilde\T_a(0,0)\tilde\T_b(0,0)\rt)}{\ltr_{\mH^{(n)}_\td}\lt(\rho^{(n)}_\td\rt)}\n
&=&\frac{\ltr_{\mH^{(n)}_{\text{I}_a}}\lt(\rhonia\,\T_a^+(x,0)\tilde\T_a(0,0)\rt)}{\ltr_{\mH^{(n)}_{\text{I}_a}}\lt(\rhonia\rt)} \frac{\ltr_{\mH^{(n)}_{\text{I}_b}}\lt(\rhonib\,\T_b^+(x,0)\tilde\T_b(0,0)\rt)}{\ltr_{\mH^{(n)}_{\text{I}_b}}\lt(\rhonib\rt)}\n
&=&\lt(\big\bra \T(x,0)\tilde\T(0,0)\big\ket_{\rho^{(n)}_{\text{I}}}\rt)^2
\eeqa
where $\mH^{(n)}_{\text{I}_a}$ and $\mH^{(n)}_{\text{I}_b}$ are the Hilbert spaces of the copies $a$ and $b$ of the $n$-copy Ising model, respectively. By comparing \eqref{mixed correlation 2T 1} and \eqref{mixed correlation 2T 2}, we see that the mixed-state two-point correlation function of the branch-point twist fields in the $n$-copy Ising model admits the representation of the form
\beqa \label{521}
&&\big\bra \T(x,0)\tilde\T(0,0)\big\ket_{\rho^{(n)}_{\text{I}}}\n
&=&\lt(\big\bra \sigma_{(1,\alpha_1)}^+(x,0)\sigma_{(1,-\alpha_1)}^+(0,0)\ket_{\rho^1_\td}\cdots \big\bra \sigma_{(n,\alpha_n)}^+(x,0)\sigma_{(n,-\alpha_n)}^+(0,0)\big\ket_{\rho^n_\td}\rt)^{1/2}
\eeqa
where the mixed-state two-point functions of $U(1)$ twist fields are known from \eqref{correlation1}.

Finally, substituting \eqref{521} into \eqref{renyi n}, we have
\beqa \label{final}
&&\hspace{-1.5cm}S_n=\frac{1}{1-n}\times \n
&&\hspace{-0.5cm}\log\lt[\varep^{\frac{1}{6}(n-\frac{1}{n})}Z_n \lt(\big\bra \sigma_{(1,\alpha_1)}^+(x,0)\sigma_{(1,-\alpha_1)}^+(0,0)
\ket_{\rho^1_\td}\cdots \big\bra \sigma_{(n,\alpha_n)}^+(x,0)\sigma_{(n,-\alpha_n)}^+(0,0)
\big\ket_{\rho^n_\td}\rt)^{\frac{1}{2}}\rt].\n
&&
\eeqa
From \eqref{final}, we can obtain the R$\acute{\text{e}}$nyi entropy for integer $n$ and its asymptotic behavior by substituting \eqref{correlation1} and using \eqref{xiangbuchulai}, respectively.

\section{Conclusion}

Generalising the program used in \cite{Ben1,Ben2,Yixiong}, we set up the Liouville space and defined the mixed-state form factor in the Dirac theory. Using the Riemann-Hilbert problem technique and low-temperature expansions, we evaluated finite-temperature form factors of $U(1)$ twist fields. Based on the structure of thermal form factors, we proposed an exact representation for form factors of $U(1)$ twist fields in general diagonal mixed states. The validity of these conjectured form factors has been confirmed by the fact that they coincide with the solution of a system of non-linear differential functional equations for mixed-state form factors of $U(1)$ twist fields. Using the Liouville-space form factor expansion,  we derived two-point correlation functions of $U(1)$ twist fields in general diagonal mixed states. Our Liouville-space method avoids re-summations of disconnected terms, which are required in the explicit calculation of the traces defining thermal averages in integrable QFT in order to cancel out divergent contributions from the partition function \cite{Pozsgay2,EssKon1,EssKon2,Pozsgay3,Sz}. These re-summations have been automatically performed in our defined mixed-state form factors and resulting form factor expansions. We deduce, from the result for the mixed-state two-point function of the   $U(1)$ twist fields, the R$\acute{\text{e}}$nyi entropy for integer $n$ in the Ising model.

In spite of the progress we have made, some open problems remain on the technical level. We have derived a recursion relation for the normalization of mixed-state form factors of $U(1)$ twist fields, but this relation hardly contribute to the evaluation of the normalization due to its $W_\nu(\theta)$-independence. So more relations involving $W_\nu(\theta)$ functions are required. It seems that the method of deducing our non-linear functional differential equations can be exploited for determining the normalization. In the present manuscript, we have not discussed applications of our method to other mixed states such as quantum quenches and non-equilibrium steady state, which in some cases will require more general $W_\nu(\theta)$ functions, including with regions of negativity and with discontinuities. With successes both in the Ising model and Dirac model, the Liouville-space method definitely deserves further investigation in more integrable models. In particular, it would be interesting to generalize our method to interacting integrable models. This is more complicated due to the fact that Wick's theorem, on which our evaluation techniques are based, is not applicable to interacting theories. Nevertheless, we can still explore functional derivatives with respect to the density matrix eigenvalues. Instead of closed equations, we would expect an infinite set of equations related to form factors
with more and more particles, which might lead to solutions or efficient expansion series.  
\vspace{1.2cm}

\noindent {\bf Acknowledgments}\\\\
I would like to thank Benjamin Doyon for his continuous support, insightful discussions, and helpful comments on the manuscript.

\appendix

\section{Proof of relation (\ref{U relation})}\label{Proof U}

Let us consider the normal-ordered operators
\beq
\Or=D^\dag_{\nu_1}(\theta_1)\cdots D^\dag_{\nu_k}(\theta_k)D_{\nu_{k+1}}(\theta_{k+1})\cdots D_{\nu_n}(\theta_n):=\prod_i D^{\ep_i}_{\nu_i}(\theta_i)
\eeq
for fixed $n$ and $k$, with all nonegative integers $n\geq k\geq0$.
Using the definition of Liouville left-action of mode operators $[D^\ep_\nu(\theta)]^\ell$ (\ref{liouville mode operators}) and of normal-ordering $\nl \cdot \nl$, we have 
\beq \label{noov}
\nl \Or^\ell \nl |\vac\ket^\rho=\prod_i \frac{\lZ^\dag_{\nu_i,\ep_i}(\theta_i)}{Q^\rho_{\nu_i,-\ep_i}(\theta_i)}|\vac\ket^\rho\,.
\eeq
By the direct calculation, we find 
\beq \label{UZU^{-1}}
\lU \lZ^\dag_{\nu,\ep}(\theta)\lU^{-1}=\lZ^\dag_{\nu,\ep}(\theta)+\ep\frac{Q^\rho_{\nu,-\ep}(\theta)}{Q^\rho_{\nu,+}(\theta)}\lZ_{\nu,-\ep}(\theta).
\eeq
Then, on the right-hand side of (\ref{U relation}), using (\ref{noov}) and (\ref{UZU^{-1}}) , we have
\beqa
\lU \nl \Or^\ell \nl |\vac\ket^\rho&=&\prod_{i=1}^{k}\lt( \frac{\lZ^\dag_{\nu_i,+}(\theta_i)}{Q^\rho_{\nu_i,-}(\theta_i)}+\frac{\lZ_{\nu_i,-}(\theta_i)}{Q^\rho_{\nu_i,+}(\theta_i)}\rt)\prod_{i=k+1}^{n}\lt( \frac{\lZ^\dag_{\nu_i,-}(\theta_i)}{Q^\rho_{\nu_i,+}(\theta_i)}-\frac{\lZ_{\nu_i,+}(\theta_i)}{Q^\rho_{\nu_i,+}(\theta_i)}\rt)|\vac\ket^\rho\no
\eeqa
and on the left-hand side, using (\ref{liouville mode operators}), we have
\beq
\Or^\ell|\vac\ket^\rho=\prod_{i=1}^{k}\lt( \frac{\lZ^\dag_{\nu_i,+}(\theta_i)}{Q^\rho_{\nu_i,-}(\theta_i)}+\frac{\lZ_{\nu_i,-}(\theta_i)}{Q^\rho_{\nu_i,+}(\theta_i)}\rt)\prod_{i=k+1}^{n}\lt( \frac{\lZ^\dag_{\nu_i,-}(\theta_i)}{Q^\rho_{\nu_i,+}(\theta_i)}+\frac{\lZ_{\nu_i,+}(\theta_i)}{Q^\rho_{\nu_i,-}(\theta_i)}\rt)|\vac\ket^\rho\no\,.
\eeq
These are equal, considering the fact that $\{\lZ^\dag_{\nu,-}(\theta), \lZ_{\nu',+}(\theta)\}=0$ and that $\lZ_{\nu,+}(\theta)|\vac\ket^\rho=0$.

\section{Recursion relation for the normalization of mixed-state form factors of $U(1)$ twist fields} \label{recursion}

In analogy to the calculation of $c_\alpha$ in \cite{James1}, a same recursion relation for the normalization $\bra \sigma^\eta_\alpha \ket_\rho$ can be obtained by considering mixed-state one-particle form factors of the fermionic primary twist fields. By similar arguments to those leading to mixed-state form factors of twist fields $\sigma^\eta_{\alpha\pm1,\alpha}$, we can also deduce mixed-state form factors  of $\sigma^\eta_{\alpha,\alpha\pm1}$. For instance, we have
\beq
{}^\rho\bra \vac| \sigma^\eta_{\alpha-1,\alpha}(0) |\theta\ket_{+,+}^\rho=
i\frac{e^{-i\pi\alpha/2}}{\Gamma(1-\alpha)} e^{2\pi i \alpha\delta_{\eta,-}} 
m^{-\alpha+1/2}e^{(\alpha-1/2)\theta} \, h^\eta_{+,+}(\theta) \bra \sigma^\eta_\alpha \ket_\rho\,.
\eeq
After a shift $\alpha \mapsto \alpha+1$, we arrive at
\beq
{}^\rho\bra \vac| \sigma^\eta_{\alpha,\alpha+1}(0) |\theta\ket_{+,+}^\rho=
-\frac{e^{-i\pi\alpha/2}}{\Gamma(-\alpha)}e^{2\pi i \alpha\delta_{\eta,-}} m^{-\alpha-1/2}e^{(\alpha+1/2)\theta}\, h^\eta_{+,+}(\theta) \bra \sigma^\eta_{\alpha+1} \ket_\rho\,.
\eeq
Notice that the leg factor $h^\eta_{+,+}(\theta) $ is invariant under the shift  $\alpha \mapsto \alpha+1$. Then, comparison with the mixed-state one-particle form factor of $\sigma^\eta_{\alpha,\alpha+1}$ given by \eqref{mixed-state ff1}
\[
{}^\rho\bra \vac| \sigma^\eta_{\alpha,\alpha+1}(0) |\theta\ket_{+,+}^\rho=
-\frac{e^{-i\pi\alpha/2}}{\Gamma(1+\alpha)}e^{2\pi i \alpha\delta_{\eta,-}} m^{\alpha+1/2}e^{(\alpha+1/2)\theta}\, h^\eta_{+,+}(\theta) \bra \sigma^\eta_{\alpha} \ket_\rho
\]
 leads to the recursion relation
\[
\frac{\bra \sigma^\eta_{\alpha+1} \ket_\rho}{\bra \sigma^\eta_{\alpha} \ket_\rho}=\frac{\Gamma(-\alpha)}{\Gamma(1+\alpha)}m^{2\alpha+1}.
\]

\end{document}